\documentclass[journal,10pt]{IEEEtran}

\makeatletter
\let\NAT@parse\undefined
\makeatother

\usepackage[english]{babel}
\usepackage{latexsym}
\usepackage{graphicx}
\usepackage{cancel}
\usepackage{cite}
\usepackage{color}
\usepackage{multirow} 
\usepackage[tight]{subfigure}

\usepackage{array}
\usepackage{mathrsfs}
\usepackage{amssymb}
\usepackage[cmex10]{amsmath}
\newtheorem{corollary}{Corollary}

\newtheorem{lemma}{Lemma}

\begin{document}
\selectlanguage{english}
\title{{The Multi-cluster Two-Wave Fading Model}}
\author{Maryam Olyaee, Juan~P.~Pe\~{n}a-Mart\'{i}n, F. J. Lopez-Martinez, and Juan~M.~Romero-Jerez

\vspace{-2mm}
\thanks{The authors are with Communications and Signal Processing Lab, Instituto Universitario de Investigaci\'on en 
Telecomunicaci\'on (TELMA), Universidad de M\'alaga, CEI Andaluc\'ia TECH.
ETSI Telecomunicaci\'on, Bulevar Louis Pasteur 35, 29010 M\'alaga, Spain. F.J. L\'opez-Mart\'inez is also with the 
Department of Signal Theory, Networking and Communications, Universidad de Granada, 18071, Granada, Spain. (e-mails: olyaee@uma.es, jppena@uma.es,  
fjlm@ugr.es, romero@dte.uma.es).
}
\thanks{This work was presented in part at the 2022 5th International Conference on Advanced Communication Technologies and Networking (CommNet)
\cite{olyaee2022}.
}
\thanks{This work was supported in part by grant PID2020-118139RB-I00 funded by MICIU/AEI/10.13039/501100011033, in part by grant P21-00420 funded by Junta de Andaluc\'ia and the European Fund for Regional Development (FEDER), and in part by grant EMERGIA20-00297 funded by Consejer\'ia de Universidad, Investigaci\'on e Innovaci\'on of Junta de Andaluc\'ia.
}
\thanks{This work has been submitted to the IEEE for publication. Copyright may
be transferred without notice, after which this version may no longer be
accessible.
}
}
\maketitle
\vspace{-10mm}

\ifCLASSOPTIONpeerreview
    \renewcommand{\baselinestretch}{2}
    \normalsize
\fi
\begin{abstract}
We introduce and characterize the Multi-cluster Two-Wave (MTW) fading model, which generalizes \textit{both} the Durgin's Two-Wave with Diffuse Power (TWDP) and the $\kappa$-$\mu$ models under a common umbrella. The MTW model consists of an arbitrary number of clusters of waves each of which may include one or two dominant (specular) components. 
The chief probability functions of the MTW fading model are obtained, including the probability density function, the cumulative distribution function and the generalized moment-generating function. \textcolor{black}{The proposed model  is empirically validated using channel measurements in the sub-THz band and} a number of applications are exemplified, including the outage probability in noise-limited and interference-limited scenarios and the energy detection probability. 
A composite Inverse Gamma (IG)/MTW  model is also investigated, thus extending the proposed propagation model to include shadowing.

\end{abstract}
\ifCLASSOPTIONjournal
\begin{IEEEkeywords}
Multicluster, Fading, Generalized Moment Generating Function, Energy Detection, Outage Probability, Composite Fading.
\end{IEEEkeywords}
\fi

%

\ifCLASSOPTIONpeerreview
    \IEEEpeerreviewmaketitle
    \renewcommand{\baselinestretch}{2}
    \normalsize
\fi
\ifCLASSOPTIONjournal
    \renewcommand{\baselinestretch}{1}
    \normalsize
\fi

\vspace{-3mm}
\section{Introduction}
\IEEEPARstart{T}he new use cases defined for 5G/6G wireless cellular communications systems, as well as the use of 
new spectral bands, including millimeter-wave (mmWave) and terahertz radio signals, give rise to new propagation scenarios not considered until now where the propagation conditions largely differ from the assumptions considered in the derivation of traditional stochastic wireless models.
Therefore, new statistical wireless fading models are necessary for the design, planning and performance evaluation of the foreseen 
wireless networks.

Rice (or Rician) fading has been traditionally used to model line-of-sight propagation scenarios, as it consists of a single cluster of waves, made up of a multitude of weakly scattered components, plus an additional constant-amplitude specular (dominant) wave with arbitrary power. The Rice fading model has been generalized in a number of ways in the literature: for instance, the $\kappa$-$\mu$ fading model was originally proposed in \cite{Yacoub01} 
and considers a signal composed of $\mu$ clusters of waves each of which may contain a specular component as in the Rice model. Within any cluster, the phases of the waves are random and have similar delay times, while the delay times among the different clusters show greater differences, so that each cluster can independently be resolved and combined at the receiver, i.e, this
physical description of the $\kappa$-$\mu$ fading implies that it can be used to model frequency-selective wireless channels typically observed in wideband propagation. The number of clusters $\mu$ should be, in principle, a positive integer; however, a better fit to experimental data is facilitated if $\mu$ is allowed to take any positive real value \cite{Yacoub07}.
It must be remarked that the mathematical description of the $\kappa$-$\mu$ model is a function, among other factors, of the total aggregate power of the specular components, but it is oblivious to the power of the individual specular components and how many clusters incorporate them. 
\textcolor{black}{Nevertheless, the fact that only one specular component per cluster is allowed limits the applicability of the $\kappa$-$\mu$ model, as experimental measurements carried out at 28 GHz \cite{rappaport2015wideband} show that more than one specular component per cluster may be expected in wideband propagation.}

On the other hand, the Two-Wave with Diffuse Power (TWDP) fading model was originally proposed in \cite{Durgin02} and provides 
an alternative generalization of the Rice model. It considers two specular components with random phases plus a diffuse 
component, representing a single cluster of multiple scattered waves. This model includes a wide variety of propagation conditions ranging from very favorable 
to worse than Rayleigh fading (hyper-Rayleigh fading) \cite{Frolik2007}. The TWDP fading model has been used to model propagation conditions in the mmWave band, and has been shown to provide a better match than the Rice fading in indoor wireless channels at 60 GHz \cite{Zochmann19}.

In this work, we introduce the Multi-cluster Two-Wave (MTW) fading model, which generalizes the Rice fading in two ways: by considering \textit{both} multiple clusters of waves (in a $\kappa$-$\mu$ fashion), and also by allowing more than one specular component within the received clusters (in a TWDP fashion). 
Thus, the proposed MTW fading model unifies both the $\kappa$-$\mu$ and TWDP models in the same mathematical framework. Encapsulating such relevant models in the same formulation permits to integrate many results previously obtained for the aforementioned models, \textcolor{black}{which have been studied until now separately and whose results are scattered in the technical literature.}
Moreover, the proposed MTW model is actually a novel fading model, as it is more general than the $\kappa$-$\mu$ and the TWDP models alone, and expands to other possible wireless environments not 
included in any of them when considered separately. 
\textcolor{black}{The MTW model is here empirically validated by field measurements and is shown to provide a better match to experimental radio propagation data than other generalized fading models in some environments.}

We derive  exact expressions for the probability density function (PDF) and the cumulative distribution function (CDF) of the received signal-to-noise ratio (SNR) under MTW fading \textcolor{black}{both in integral form and in series form.} 
Additionally, the generalized moment generating function (GMGF) is given in closed-form, which allows to obtain many different performance metrics of wireless communications systems undergoing the new proposed model. In particular, closed-form expressions are obtained for the outage probability in interference-limited and noise-limited scenarios, and for the energy detection probability. We also exemplify how the MTW model can be further generalized to include the effect of shadowing, in a straightforward way \cite{ramirez2021composite}.


The rest of this paper is organized as follows: In Section II, the MTW channel model is presented, and the PDF, CDF, and generalized MGF are derived. \textcolor{black}{In Section III, an empirical validation of the MTW model is presented.} In Section IV, expressions for different performance metrics for the proposed model are obtained, including the outage probability, the energy detection probability and composite fading modeling.
Numerical results are given in Section V, followed by the concluding remarks in Section VI.

\section{MTW channel model} \label{math1} 
The Multi-cluster Two-Wave (MTW) fading model results from the combination of multiple signal clusters, being \textcolor{black}{any of them accompanied by up to two} specular components representing dominant waves. 
Let us assume that $\mu$ signal clusters are received. The complex baseband signal amplitude of the \emph{i-th} cluster can be expressed as
\begin{equation} 	\label{eq:001} 	
Z_i  =  V_{i,1}  \exp \left( {j\phi _{i,1} } \right) +  V_{i,2}  \exp \left( {j\phi _{i,2} } \right) + X_i + jY_i, 
\end{equation} 
where $ V_{i,k} $ and $\phi _{i,k}$ indicate, respectively, the 
\emph{k-th} specular component $(k=1,2)$ amplitude and the uniformly distributed random phase in the interval $[0, 2\pi)$,  
$\phi _{i,k}  \sim \mathcal{U}[0,2\pi)$.
On the other hand, $(X_i + jY_i)$ is a complex Gaussian random variable (RV) with 
$X_i,Y_i \sim \mathcal{N}(0,\sigma^2)$, representing the contributions due to the combined reception of numerous weak  
scattered waves within the \emph{i-th} cluster. 

In the MTW fading model, {\textcolor{black}{the received waves within a given cluster have very small delays (negligible, in practice) among them as well as arbitrary phases.
On the other hand, the delay-time spreads of the different clusters are assumed to be large enough so that they can be individually detected and combined at the receiver, in such a way that the 
received power is a result of the 
summation of the different clusters powers, i.e., a wideband propagation scenario is inherently assumed.} Thus, we can write $W = R^2=\sum_{i=1}^\mu { |Z_i|^2}$, where $W$ and $R$ represent
the total instantaneous received power and signal envelope, respectively.


\subsection{Probability distribution (PDF and CDF)}
In the following, the PDF of the SNR of the MTW fading model is derived.
Let us define the following RVs:
\begin{equation}
	\label{eq:003}
	\begin{split}
		\theta_i & \triangleq \phi _{i,1} - \phi _{i,2}
		\\
		d_{\theta}^2 & \triangleq \sum\limits_{i = 1}^\mu \left(V_{i,1} \cos\phi _{i,1} + V_{1,2}\cos \phi _{i,2}  \right)^2
		\\& + \left(V_{i,1} \sin \phi _{i,1} + V_{i,2}\sin \phi _{i,2}  \right)^2
		\\&
		=\sum\limits_{i = 1}^\mu \left(V_{i,1}^2 + V_{i,2}^2 + 2V_{i,1} V_{i,2} \cos\theta_i\right).
	\end{split}
\end{equation}
Note that $d_{\theta}^2$ represents the aggregated power of the specular components in terms of the RVs $\theta_i$,
which follow a triangular distribution in the interval $ [- 2\pi, 2\pi) $, and which are equivalent to a uniform 
distribution in $[0,2\pi)$ due to the $2\pi$ periodicity of the phases. 

The signal complex amplitude of the \emph{i-th} cluster given in (\ref{eq:001}) can be rewritten in terms of $\theta_i$ as
\begin{equation}
	\label{eq:004}
	Z_i  = e^{j\phi _{i,2}} \left(V_{i,1}  e^ {j\theta_i }  +  V_{i,2} \right)  + X_i + jY_i. 
\end{equation}
Thus, for a given particular realization of variable $\theta_i$, (\ref{eq:004}) is equivalent to the signal of a cluster 
with a single specular component with amplitude $\left|{V_{i,1}  \exp \left( {j\theta_i } \right) +  V_{i,2} } \right|$.
Therefore, the set of clusters forms a $\kappa$-$\mu$ channel as in \cite{Yacoub07} when conditioned on {\textcolor{black}{$\theta \triangleq \{\theta_1,\theta_2,\ldots,\theta_\mu \}$}. Hence, the conditional PDF of the received signal power for a given $\theta$, 
$W_\theta$, will be expressed as
\begin{equation}
	\label{eq:006}
	\begin{split}
		f_{W_\theta}(w) & = \frac{1}{2\sigma^2} \left(\frac{w}{d_\theta^2} \right)^{\frac{\mu-1}{2}} \exp \left(-\frac{w}{2\sigma^2}
-\frac{d_\theta^2}{2\sigma^2} \right)
		\\ & \times
		I_{\mu-1}\left(\sqrt{w}\frac{d_\theta}{\sigma^2} \right),
	\end{split}
\end{equation}
where $I_{\nu}$ is the modified Bessel function of the first kind with order $\nu$.
Introducing the parameter
\vspace{-2mm}
\begin{equation}
	\label{eq:007}
	\kappa_\theta \triangleq \frac{d_\theta ^2 }{2\sigma^2 \mu},
\end{equation}
which represents the ratio between the specular components power, conditioned on $\theta$, and the total average diffuse power 
from all the clusters, we can write the average 
conditioned power as
\begin{equation}
	\label{eq:010}
	\overline{W_{\theta}} = d_{\theta}^2  + 2\sigma^2 \mu = 2\sigma^2 \mu (1+\kappa_\theta).
\end{equation}
Thus, by combining  (\ref{eq:007}) and (\ref{eq:010}), we can write
\begin{equation}
	\label{eq:011}
	\begin{split}
		d_\theta ^2 &= \frac{\overline{W_\theta} }{(1+\kappa_\theta )}  \kappa_\theta.
	\end{split}
\end{equation}
Let us now introduce the random variable $\gamma \triangleq W E_s / N_0 $ representing the received SNR, where $E_s$ is the 
symbol energy and $N_0$  is the one-sided AWGN power spectral density. Thus, denoting by $\gamma_\theta$ and $\overline{\gamma_\theta}$, respectively,
the conditioned instantaneous and average SNR, we have that $E_s / N_0 =  \frac{\overline{\gamma_\theta}}{\overline{W_\theta}} $, and
\begin{equation}
	\label{eq:012}
	\gamma_{\theta} = W_\theta \frac{\overline{\gamma_\theta}}{\overline{W_\theta}}.
\end{equation}
Thus, from \eqref{eq:006} and considering (\ref{eq:007})-(\ref{eq:012}), the PDF of the conditioned instantaneous SNR will be
\begin{equation}
	\label{eq:013}
	\begin{split}
		f_{\gamma_\theta}(x) & =  \frac{\overline{W_\theta}}{\overline{\gamma_\theta}}  f_{W_\theta} \left(\frac{\overline{W_\theta
}}{\overline{\gamma_\theta}} x  \right)
		\\&
		= \frac{{\mu {{\left( {1 + {\kappa _\theta }} \right)}^{\frac{{\mu  + 1}}{2}}}}}{{\overline {{\gamma _\theta }} {e^{\mu {
\kappa _\theta }}}\;}}{\left( {\frac{x}{{\overline {{\gamma _\theta }} {\kappa _\theta }}}} \right)^{\frac{{\mu  - 1}}{2}}}{e^{ 
- \frac{{\mu \left( {1 + {\kappa _\theta }} \right)}}{{\overline {{\gamma _\theta }} }}x}}
		\\& \times
		{I_{\mu  - 1}}\left( {2\mu \sqrt {\frac{{{\kappa _\theta }\left( {1 + {\kappa _\theta }} \right)}}{{\overline {{\gamma _
\theta }} \;}}x} } \right).
	\end{split}
\end{equation}
As the expected unconditional average power is $\overline W  =  \sum\limits_{j = 1}^\mu \left(V_{j,1}^2 + V_{j,2}^2 \right) + 2{\sigma ^2}\mu$, from (\ref{eq:012}) and  (\ref{eq:010}) we can write
\begin{equation}
	\label{eq:015a}
	\begin{split}
		\overline {{\gamma _\theta }} &
		= \overline \gamma  \frac{{\overline {{W_\theta }} \;}}{{\overline W }}
		\\&
		= \overline \gamma  \frac{\sum\limits_{i = 1}^\mu \left(V_{i,1}^2 + V_{i,2}^2  + 2V_{i,1}V_{i,2}\cos \theta_i\right) + 2{\sigma ^2}\mu \;}
		{{\sum\limits_{j = 1}^\mu \left(V_{j,1}^2 + V_{j,2}^2\right)  + 2{\sigma ^2}\mu }}
		\\&
		= \overline \gamma  \left( {1 + \frac{{\sum\limits_{i = 1}^\mu 2V_{i,1}}{V_{i,2}\cos \theta_i \;}}{{\sum\limits_{j = 1}^\mu \left(V_{j,1}^2 + V_{j,2}^2 \right) + 2{\sigma ^2}\mu}}} \right),
	\end{split}
\end{equation}
The MTW fading model can be conveniently described in terms of parameters $K$ and $\Delta_i$,  defined as
{\textcolor{black}{
\begin{equation}
	\label{eq:014}
	\begin{split}
		& K \triangleq \frac{{\sum\limits_{i = 1}^\mu \left(V_{i,1}^2 + V_{i,2}^2 \right)}}{{2{\sigma ^2}\mu }},
		\\&
		\Delta_i  \triangleq \frac{{2{V_{i,1}}{V_{i,2}}}}{{\sum\limits_{j = 1}^\mu \left(V_{j,1}^2 + V_{j,2}^2\right)}},
	\end{split}
\end{equation}}
where $K$ represents the ratio between the average power of the specular components and the diffuse components power from all the clusters; 
and $\Delta_i$ provides a measure of the asymmetry of the specular components of the \emph{i-th} cluster, verifying $0\leq \Delta_i \leq \left(V_{i,1}^2 + V_{i,2}^2\right) / \left(\sum\limits_{j = 1}^\mu \left(V_{j,1}^2 + V_{j,2}^2\right)\right)$, {\textcolor{black}{which implies 
the inequalities $0\leq \sum\limits_{j = 1}^\mu \Delta_i \leq 1$.
The more similar the amplitude of the specular components of the \emph{i-th} cluster the higher the value of $\Delta_i$, and $\Delta_i=0$ is attained when only one specular component, if any, is present in the cluster, i.e. $V_{i,1}=0$ and/or $V_{i,2}=0$.} 
When all the specular power is concentrated in the \emph{i-th} cluster or it is the only one received, then $ \Delta_i =1 $ can be reached, indicating that both specular components of that cluster have the same amplitude. 
{\textcolor{black}{In practical wireless environments not all the received wave clusters must necessarily contain two specular components, as some of them may contain only one or even none. Let $N$, with $0 \leq N \leq \mu$, be the number of clusters containing two specular components, which will be denoted as clusters $1,\ldots,N$. The remaining $\mu -N$ clusters will have one or none specular components, verifying $\Delta_j=0$ for $j=N+1,\ldots,\mu$.
In this regard, note that the statistical characterization of the SNR is oblivious of the order of arrival of the wave clusters and, therefore, of its numeration, which is only established for the sake of clarity of the presentation of the model.
Then, after some manipulations, (\ref{eq:015a}) can be written in terms of parameters $K$ and $\Delta_i$ as}
\begin{equation}
	\label{eq:015b}
	\begin{split}
		\overline {{\gamma _\theta }} &
		= \overline \gamma  \frac{{1 + K\left( {1 + \sum\limits_{i = 1}^N \Delta_i \cos \theta_i } \right)}}{{1 + K}},
	\end{split}
\end{equation}
and with the help of (\ref{eq:003}), (\ref{eq:007}) and (\ref{eq:014}) we can write
\begin{equation}
	\label{eq:016}
	{\kappa _\theta }  = K\left( {1 + \sum\limits_{i = 1}^N \Delta_i \cos \theta_i } \right).
\end{equation}
Therefore, from (\ref{eq:015b}) and (\ref{eq:016}) we conclude that 
\begin{equation}
	\label{eq:017}
	\frac{{1 + {\kappa _\theta }}}{{\overline {{\gamma _\theta }} \;}} =
	\frac{{1 + K\left( {1 + \sum\limits_{i = 1}^N \Delta_i \cos \theta_i } \right)}}{{\overline \gamma  \frac{{1 + K\left( {1 + \sum\limits_{i = 1}^N \Delta_i \cos \theta_i } \right)
}}{{1 + K}}\;}}
	= \frac{1 + K}{\overline \gamma },
\end{equation}
which shows that the ratio $(1 + \kappa _\theta) / \overline {\gamma _\theta }$ is invariant with respect to {\textcolor{black}{$\theta_1,\ldots,\theta_N$}. 
Substituting (\ref{eq:017}) 
into (\ref{eq:013}), the conditional PDF of the SNR becomes
\begin{equation}
	\label{eq:018}
	\begin{split}
		f_{\gamma_\theta}(x) &=
		\frac{ \mu {\left( {1 + K} \right)}^{\frac{\mu  + 1}{2}}}
		{\overline \gamma  e^{\mu \kappa _\theta}}
		\left( {\frac{x}{{\overline \gamma  \kappa _\theta}}} \right)^{\frac{{\mu  - 1}}{2}}
		\\& \times
		{e^{ - \frac{{\mu \left( {1 + K} \right)}}{{\overline \gamma  }}x}}
		I_{\mu  - 1}\left( {2\mu \sqrt {\frac{{\kappa _\theta\left( {1 + K} \right)}}{{\overline \gamma }}x} } \right).
	\end{split}
\end{equation}

 The unconditional PDF of the SNR $\gamma$ can be obtained by averaging over the realizations of the RVs \textcolor{black}{$\theta_1,\ldots,\theta_N$}, which yields the result given in the following lemma.
\begin{lemma}
	\label{teorema1}
	Let $\gamma$ represent the received SNR in an MTW fading channel. Then, the PDF of $\gamma$ is given by (\ref{eq:019}).
\begin{figure*}[!t]
\normalsize
	\begin{equation}
		\label{eq:019}
		\begin{split}
			f_{\gamma }(x) =&
			\frac{\mu}{\pi^N} \left(\frac{1+K}{\overline \gamma}\right)^{\frac{\mu  + 1}{2}} e^{-\mu K}
			\left(\frac{x}{K}\right)^{\frac{\mu  - 1}{2}}  e^{ - \mu\frac{ 1 + K}{\overline \gamma  }x}
			\int_{ \theta_1=0 }^\pi \cdots \int_{ \theta_N=0 }^\pi {e^{ - \mu K \sum\limits_{i = 1}^N \Delta_i \cos \theta_i } \left(1+\sum\limits_{i = 1}^N \Delta_i \cos \theta_i \right)^{\frac{{1-\mu}}{2}}}
			\\& \times
			I_{\mu  - 1}\left( 2\mu \sqrt {\frac{ 1 + K}{\overline \gamma } K\left(1+\sum\limits_{i = 1}^N \Delta_i \cos \theta_i\right) x}  \right) d\theta_1 \cdots d\theta_N.
		\end{split}
	\end{equation}
\hrulefill
\vspace*{4pt}
\end{figure*}
\end{lemma}
\begin{IEEEproof}
	Taking into account that $\cos \theta_i$ is symmetric around $\pi$, and the fact that the PDF of $\theta_i$ is $f_{\theta_i} (\theta_i)=\frac{1}{2\pi}$, constant in the interval $[0,2\pi)$,  (\ref{eq:019}) is obtained  by considering (\ref{eq:016}) and averaging (\ref{eq:018}) with respect to {\textcolor{black}{$\theta_1,\ldots,\theta_N$}}.
\end{IEEEproof}

For $N=\mu=1$, the MTW model collapses to the TWDP one, and (\ref{eq:019}) 
is equivalent to \cite[eq. (26)]{Rao15}. On the other hand, if $\Delta_i=0$, (\ref{eq:019}) is equivalent to \cite[eq. (2)]{Yacoub07} and 
we obtain the $\kappa$-$\mu$ channel model (note that in this case the integrations are trivial and can be solved in closed-form). To the best of our knowledge, no other previously proposed
statistical fading model contains such different models such as $\kappa$-$\mu$ and TWDP as special cases, which gives a great versatility to the MTW fading model. 

\begin{corollary}
	\label{corolario3}
	The CDF of the received SNR $\gamma$ in an MTW fading channel is given by
	\begin{equation}
		\label{eq:019d}
		\begin{split}
		&F_{\gamma }(x) = 1-\frac{1}{\pi^N} \int_{ \theta_1=0 }^\pi \cdots \int_{ \theta_N=0 }^\pi \\& 
		 {Q_\mu \left( \sqrt{2\mu K\left(1+\sum\limits_{i = 1}^N \Delta_i \cos \theta_i\right)}, \sqrt{2x \mu \zeta} \right)d\theta_1 \cdots d\theta_N},
		\end{split}
	\end{equation}
		with $\zeta \triangleq \frac{1+K}{\overline \gamma}$ and
where $Q_\nu (a,b)$ is the $\nu$-th order generalized Marcum $Q$-function as defined as \cite[eq. (4.60)]{Simon05}
	\begin{equation}
		\label{eq:019cc}
		Q_\nu (a,b)= a^{1-\nu}\int_b^\infty x^\nu \exp\left[- \frac{x^2+a^2}{2} \right] I_{\nu-1}(ax)dx.
	\end{equation}
\end{corollary}
\begin{IEEEproof}
	Considering
	\begin{equation} \label{eq:019bis}
		F_{\gamma }(x)  = \int_0^x f_{\gamma }(t) dt
		= 1-\int_x^\infty f_{\gamma }(t) dt,
	\end{equation}
 	and introducing  (\ref{eq:019}) into (\ref{eq:019bis}), the result given in (\ref{eq:019d}) is obtained, after some 
manipulation, by changing the integration order of
 variables {\textcolor{black}{$\theta_1,\ldots,\theta_N$}} and $t$ and performing the 
transformation $t=\frac{z^2}{2\mu \zeta}$.
\end{IEEEproof}

\textcolor{black}{The expressions given in (\ref {eq:019}) and (\ref {eq:019d}) for, respectively, the PDF and CDF, can not strictly be considered closed-form, as they are given in terms of definite integrations. However the integrands in both cases are bounded continuous functions and therefore the integrals can be calculated without difficulty. Numerical computation of finite-range integrals of smooth functions are common in communication theory \cite{Simon05} and do not represent a computational challenge. 
It is worth mentioning that the number $N$ of clusters with two specular components will not typically be higher than 3 (see for example \cite{rappaport2015wideband}), which simplifies the computation of the nested integrals, as 2 and 3 nested integrations are in-built functions in widely used computational packages. Moreover, the simpler case $N=1$ captures most of the essence of the MTW distribution and is able to provide a better fit to empirical measurements than other generalized fading models, as it will be shown in Section III. 
Nevertheless, the obtained integral PDF and CDF expressions may be difficult to manipulate in subsequent performance analysis derivations. Fortunately, much easy-to-manipulate expressions can be obtained for these functions in terms of series, as we show in the following lemma.}

\textcolor{black}{
\begin{lemma}
	\label{teorema3}
	Let $\gamma$ represent the received SNR in an MTW fading channel. Then, the PDF and CDF of $\gamma$ can be written, respectively,  in terms of series as 
	\begin{equation} \label{eq:501}
		f_\gamma  (x) = e^{ - \beta x} \sum\limits_{k = 0}^\infty  {X\left( k \right)} x^{\mu  + k - 1},
	\end{equation}
	\begin{equation} \label{eq:502}
		F_\gamma  (x) = \sum\limits_{k = 0}^\infty  {X\left( k \right)\beta ^{ - \mu  - k} } \gamma \left( {\mu  + k,\beta x} \right),
	\end{equation}
	where $X\left( k \right)$ is given in (\ref{eq:503}) and $\gamma \left( {\cdot,\cdot} \right)$ is the lower incomplete Gamma function  \cite[(8.350)]{Gradshteyn00}, and where $\tau(r,N)$ in (\ref{eq:503}) is defined as the set of $N$-tuples such that ${\tau(r,N)=\left\{(r_1,r_2,\cdots,r_N):r_i \in \mathbb{N},\sum_{i=1}^N r_i=r \right\}}$.
	\begin{figure*}[!t]
\normalsize
\textcolor{black}{
\begin{equation}
\begin{split}
\label{eq:503}
X(k) = e^{ - \mu K} K^k \mu ^{\mu  + 2k} \left( {\frac{{1 + K}}
{{\overline \gamma  }}} \right)^{\mu  + k} \frac{1}
{{k!\Gamma \left( {\mu  + k} \right)}}\sum\limits_{r = 0}^k \binom{k}{r} r!
\sum\limits_{\tau (r,N)} \prod \limits_{i = 1}^N \left[{\frac{1}{{r_i !}}
\left(\frac{{\Delta _i }}{{2}}\right)^{r_i }
\sum\limits_{l = 0}^{r_i } \binom{r_i}{l} I_{2l - r_i } \left( { - \mu K\Delta _i } \right)} \right] .
\end{split}
\end{equation}
}
\hrulefill
\vspace*{4pt}
\end{figure*}
\end{lemma}
\begin{IEEEproof}
See Appendix \ref{apendiceA}.
\end{IEEEproof}}

\textcolor{black}{
Note that Lemma \ref{teorema3} states that the MTW distribution can be expressed as a mixture of Gamma distributions, which permits to readily obtain several relevant performance metrics \cite{atapattu2011mixture}.
Moreover, as we show in the next subsection, the GMGF of the MTW model can be expressed in closed-form in terms of a finite number of well-known functions, which permits to directly obtain different performance metrics, as well as the MGF and the moments of the distribution, also in closed-form circumventing numerical integrations and infinite summations.}

\subsection{Generalized moment generating function}
The generalized MGF of the SNR is defined as \cite{Pena17b}
\begin{equation}
	\label{eq:022}
	\phi _\gamma ^{(n)} \left( s \right) \triangleq  E\left\{ {\gamma ^n e^{\gamma s} } \right\} = \int_0^\infty  {x^n e^{xs} f_
\gamma  } \left( x \right)dx,
\end{equation}
where $E\left\{\cdot\right\}$ denotes the expectation operator. 
In the sequel, we will assume $n\in \mathbb{N}$. 

\begin{lemma}
\label{teorema2}
Let $\gamma$ represent the received SNR in an MTW fading channel.
Then,  the generalized MGF of $\gamma$, $\phi_\gamma ^{(n)} (s)$, is given by (\ref{eq:023}), where $\tau(r,N)$ is defined as in Lemma \ref{teorema3}.
\end{lemma}
\begin{IEEEproof}
See Appendix \ref{apendiceB}.
\end{IEEEproof}
\begin{figure*}[!t]
\normalsize
\begin{equation}
\begin{split}
\label{eq:023}
  \phi _\gamma ^{(n)} \left( s \right) =& \overline \gamma  ^n \Gamma \left( {\mu  + n} \right)\exp \left( {\frac{{\mu K\overline \gamma  s}}
{{\mu \left( {1 + K} \right) - \overline \gamma  s}}} \right)\sum\limits_{q = 0}^n  \binom{n}{q} \frac{{\left( {\mu K} \right)^q }}
{{\Gamma \left( {\mu  + q} \right)}}\frac{{\left( {\mu \left( {1 + K} \right)} \right)^{\mu  + q} }}
{{\left( {\mu \left( {1 + K} \right) - \overline \gamma  s} \right)^{\mu  + q + n} }} \hfill \\&
\sum\limits_{r = 0}^q  \binom{q}{r} r!\sum\limits_{\tau (r,N)} {\prod\limits_{i = 1}^N {\left[ {\frac{1}
{{r_i !}}\left( {\frac{{\Delta _i }}
{2}} \right)^{r_i } \sum\limits_{l = 0}^{r_i }  \binom{r_i}{l} I_{2l - r_i } \left( {\frac{{\mu K\Delta _i \overline \gamma  s}}
{{\mu \left( {1 + K} \right) - \overline \gamma  s}}} \right)} \right]} }. \hfill \\ 
\end{split}
\end{equation}
\hrulefill
\vspace*{4pt}
\end{figure*}

The MGF and all the moments of the MTW distribution can be readily obtained from the GMGF as we show in the following.

\begin{corollary}
\label{corolario33}
The MGF of the SNR $\gamma$ in an MTW fading channel is given by
\begin{equation}
\begin{split}
\label{eq:108}
{M_\gamma }(s)  = &
\left( {\frac{{\mu \left( {1 + K} \right)}}
{{\mu \left( {1 + K} \right) - \overline \gamma  s}}} \right)^\mu  
  \exp \left( {\frac{{\mu K\overline \gamma  s}}
{{\mu \left( {1 + K} \right) - \overline \gamma  s}}} \right) \\&  \times 
\prod\limits_{i = 1}^N  {I_0 \left( {\frac{{\mu K\Delta _i \overline \gamma  s}}
{{\mu \left( {1 + K} \right) - \overline \gamma  s}}} \right)}.
\end{split}
\end{equation}
\end{corollary}
\begin{IEEEproof}
The MGF is defined as ${M_\gamma }(s) \triangleq E\left\{ {e^{\gamma s} } \right\} = \phi _\gamma ^{(0)} \left( s \right)$.
Thus, substituting $n=0$ in  (\ref{eq:023}), (\ref{eq:108}) is obtained.
\end{IEEEproof}

\textcolor{black}{It is worth noting that a closed-form expression of the MGF permits to readily obtain the bit/symbol error rate for many modulation schemes when a multi-antenna receiver employing Maximal Ratio Combining (MRC) is used at the receiver \cite{Simon05}.}

\begin{corollary}
\label{corolario4}
The $n$-th order non-central moment of  the SNR $\gamma$ is given by
\begin{equation}
\begin{split}
\label{eq:109}
  \nu _n  &= \left( {\frac{{\overline \gamma  }}
{{\mu \left( {1 + K} \right)}}} \right)^n \Gamma \left( {\mu  + n} \right)\sum\limits_{q = 0}^n  \binom{n}{q} \frac{{\left( {\mu K} \right)^q }}
{{\Gamma \left( {\mu  + q} \right)}} \hfill \\&
\sum\limits_{r = 0}^q  \binom{q}{r} r!\sum\limits_{\tau (r,N)} {\prod\limits_{i = 1}^N {\left[ {\frac{1}
{{r_i !}}\left( {\frac{{\Delta _i }}
{2}} \right)^{r_i } \sum\limits_{l = 0}^{r_i }  \binom{r_i}{l} \delta _{2l,r_i } } \right]} } ,
\end{split}
\end{equation}
where $\delta_{2l,m}$ is the kronecker delta function, \textcolor{black}{with $\delta_{i,j}=1$ for $i=j$ and $0$ otherwise \cite{Gradshteyn00}}.
\begin{IEEEproof}
The proof follows from (\ref{eq:023}) by considering that  $\nu _n  = \phi _\gamma ^{(n)} \left( 0 \right)$ and noting that 
$I_0(0)=1$ and $I_{\nu}(0)=0$ for $\nu \neq 0$.
\end{IEEEproof}
\end{corollary}

It is easy to check that $\nu _0  =1$ and $\nu _1  = \overline{\gamma}$, as expected. 

The amount of fading (AoF) is a fading metric introduced in \cite{Charash79} as the SNR variance normalized to its squared mean, i.e.,
\begin{equation}
\label{eq:110}
\text{AoF}=\frac{{\rm Var}(\gamma)}{({\rm E}[\gamma])^2}= \frac{\nu _2 - \overline{\gamma}^2}{\overline{\gamma}^2}.
\end{equation}
The AoF for the MTW can be calculated as stated in the next corollary.

\begin{corollary}
\label{corolario5}
The AoF in an MTW fading channel is given by
\begin{equation}
\label{eq:111}
  \text{AoF} =\frac{1}
{{\left( {1 + K} \right)^2 }}\left[ {\frac{{1 + 2K}}
{\mu } + K^2 \left( {\sum\limits_{i = 1}^N {\frac{{\Delta _i^2 }}
{2}} } \right)} \right].
\end{equation}
\end{corollary}
\begin{IEEEproof}
The proof follows from (\ref{eq:109}) and (\ref{eq:110}) by simple substitution.
\end{IEEEproof}


\subsection{Asymptotic CDF and diversity order}

An asymptotic expression of the CDF in the high SNR regime is useful to obtain insight of the effect of system parameters on performance,
as the CDF is closely related to the outage probability, as we will later show. To obtain such expression we first note that, by
taking the limit $s\rightarrow\infty$ in the MGF given in (\ref{eq:108}), the following expression holds:
\begin{equation}
\label{eq:112}
\begin{split}
\left| {M_\gamma  (s)} \right| = &
\frac{{\mu ^\mu  \left( {1 + K} \right)^\mu  }}
{{e^{\mu K} }}\left[ {\prod\limits_{i = 1}^N  {I_0 \left( {\mu K\Delta _i } \right)} } \right] \\& \times \frac{1}
{{\overline \gamma  ^\mu  \left| s \right|^\mu  }} + o\left( {\left| s \right|^{ - \mu } } \right),
\end{split}
\end{equation}
where $o(x)$ represents a function $g(x)$ verifying $\lim_{x\rightarrow 0} g(x)/x=0.$
Then, from Propositions 3 and 5 in \cite{Wang03} the asymptotic CDF can be expressed as
\begin{equation}
\label{eq:113}
F_\gamma  (x) =
\frac{{\mu ^\mu  \left( {1 + K} \right)^\mu  }}
{{\Gamma \left( {\mu  + 1} \right)e^{\mu K} }}\left[ {\prod\limits_{i = 1}^N  {I_0 \left( {\mu K\Delta _i } \right)} } \right]\left( {\frac{x}
{{\overline \gamma  }}} \right)^\mu   + o\left( {\overline \gamma  ^{ - \mu } } \right).
\end{equation}
It is clear from (\ref{eq:113}) that the diversity order of the MTW model is $\mu$.

{
\section{\textcolor{black}{Empirical Validation}}\label{secVal}
\textcolor{black}{Before moving into the performance analysis evaluation, we first wish to validate the adequacy of the MTW model with field measurements. Specifically, we use a subset of the experimental data used in \cite{fitting1} for an outdoor deployment in the sub-THz band at 142 GHz, as described in \cite{fitting2}. For exemplary purposes, we considered the scenarios given in \cite[Figs. 2a/2e/2f]{fitting1}, respectively denoted as A, E and F in the sequel. 
To facilitate the validation and for the fairness of the comparison with other models, we will consider the case $N=1$ in the MTW model, denoting $\Delta\triangleq \Delta_1$. In this case, the MTW model is restricted to 3 channel parameters, namely $\Delta$, $K$ and $\mu$.
Our goal is to test the efficacy of the MTW distribution against two popular reference models with also 3 parameters used for benchmarking: the Fluctuating Two-Ray (FTR) \cite{Romero17} and the $\kappa$-$\mu$ \cite{Paris14} Shadowed (KMS) fading models. To measure the goodness of fit, we use the mean-square-error (MSE) test as a well-established metric to capture fitness accuracy, given by the following expression:
\begin{align}
\label{eqmse}
\text{MSE}=\frac{1}{T}\sum_{i=1}^{T}\left ( \widehat{f}_R(r_i)-f_R(r_i)\right )^{2},
\end{align}
where $\widehat{f}_R(\cdot)$ denotes the empirical PDF of the signal envelope and $f_R(\cdot)$ denotes the theoretical PDF which, by a simple change of variables, can be calculated from either (\ref{eq:019}) or (\ref{eq:501}) as $f_r(r)=2r f_{\gamma}(r^2)$. On the other hand, $T$ is the sample size of the empirical PDF.}
\begin{table*}[t]
\caption{{Fitting Results of MTW, FTR and $\kappa$-$\mu$ shadowed fading models for different scenarios.}} 
\label{Tablefit}
\centering
\begin{tabular}{|c|c|c|c|c|c|c|} 
\hline
\textbf{Case}               & \textbf{Distribution}   & $K$   & $\Delta$ & $\mu$ & $m$   & \textbf{MSE}      \\ 
\hline
\multirow{3}{*}{\textbf{A}} & MTW                     & 29.63 & 0.28     & 8.17  & -     & \textbf{0.00398}  \\ 
\cline{2-7}
                            & FTR                     & 38.14 & 0.003    & -     & 79.32 & 0.32746           \\ 
\cline{2-7}
                            & KMS										  & 11.10 & -        & 2.99  & 89.34 & 0.32772           \\ 
\hline
\multirow{3}{*}{\textbf{E}} & MTW                     & 25.06 & 0.40     & 1.89  & -     & \textbf{0.00111}  \\ 
\cline{2-7}
                            & FTR                     & 67.29 & 0.39     & -     & 75.19 & 0.00276           \\ 
\cline{2-7}
                            & KMS										  & 24.83 & 0        & 0.60  & 73.77 & 0.02689           \\ 
\hline
\multirow{3}{*}{\textbf{F}} & MTW                     & 11.38 & 0.61     & 0.57  & -     & \textbf{0.00346}  \\ 
\cline{2-7}
                            & FTR                     & 6.20  & 0.66     & 1     & 95.61 & 0.00362           \\ 
\cline{2-7}
                            & KMS										  & 39.99 & -        & 0.107 & 78.74 & 0.00392           \\
\hline
\end{tabular}
\end{table*}
{
\textcolor{black}{In Table~\ref{Tablefit}, the corresponding values for the MSE test are presented, together with the estimated parameters for each of the candidate fading models. The optimization was carried out using MATLAB's \emph{fminsearch} function. Bold-faced fonts are used to denote the best-fitting alternative for each scenario.}

\textcolor{black}{It can be observed that, according to the MSE criteria, the MTW fading  model provides the best fitting results in all the considered scenarios, and the improvement of the MTW model with respect to the other compared ones is remarkable in case A. Note that, as in the KMS model, parameter $\mu$, although originally defined as an integer, is allowed to take any positive real value to improve the fitting results. In Fig.~\ref{fig:fitting}, we compare the theoretical PDFs of the MTW, FTR, and KMS fading models against the sub-THz channel measurements for the environments described in~\cite[Fig.~1]{fitting1}. It can be seen that the MTW model yields a more accurate fit to the empirical distributions in all cases, which agrees with the MSE values shown in Table~\ref{Tablefit}. These results show that, for some channel measurements, the proposed MTW fading model offers a good balance between versatility, complexity, goodness of fit and physically-motivated modeling. For instance, for the environment in Fig.~\ref{fig:fitting}a (Case A), the MTW captures the richness of the channel scattering through parameter $\mu$ (i.e., $8.17$), which represents the diversity order. This parameter controls the appearance of a bimodal behavior in the PDF, thereby matching the empirical measurements. Moreover, parameter $\mu$  also controls the slope of the PDF and CDF for low values of the signal envelope, which also permits to fit the empirical data as they move away from the origin on the x-axis. In particular, this behavior is present in Figs.~\ref{fig:fitting}a (Case A) and~\ref{fig:fitting}b (Case E), where the empirical distributions start at approximately $0.7$ and $0.5$ on the x-axis, respectively. Conversely, in Fig.~\ref{fig:fitting}c (Case F), the empirical measurements begin at $0$ on the x-axis, which translates to small values of $\mu$. It is worth mentioning that the FTR model can potentially present a bimodal behavior, but its diversity order is always 1 and, therefore, it cannot properly adapt to a rich scattering environment and the shift of the data on the x-axis. On the other hand, although the KMS also has a diversity order $\mu$, it cannot fit a bimodal behavior as every cluster may only have one specular component.
}

\begin{figure*}[ht!]
    \centering
    \subfigure[Case A] {\includegraphics[width=0.32\textwidth]{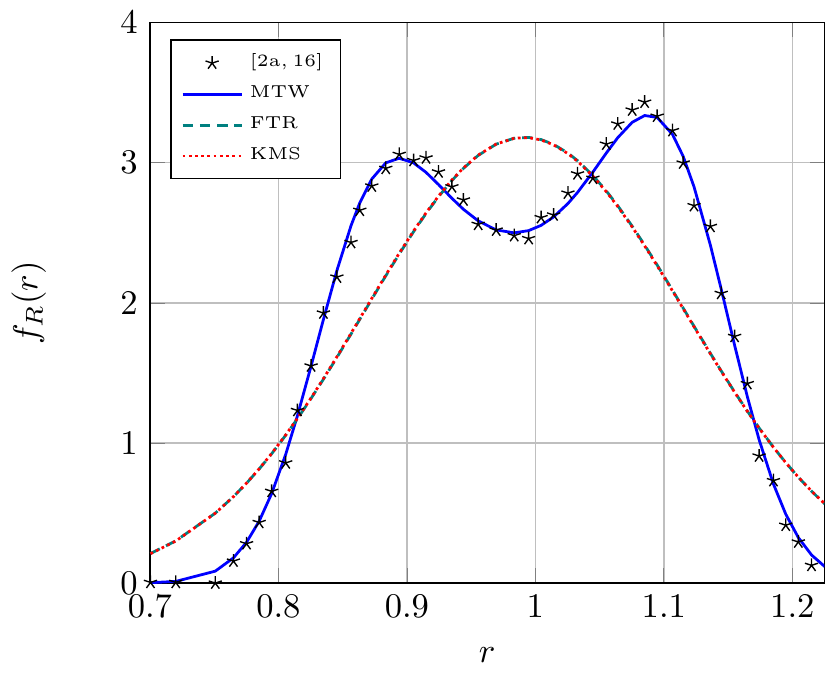}} 
    \subfigure[Case E]{\includegraphics[width=0.32\textwidth]{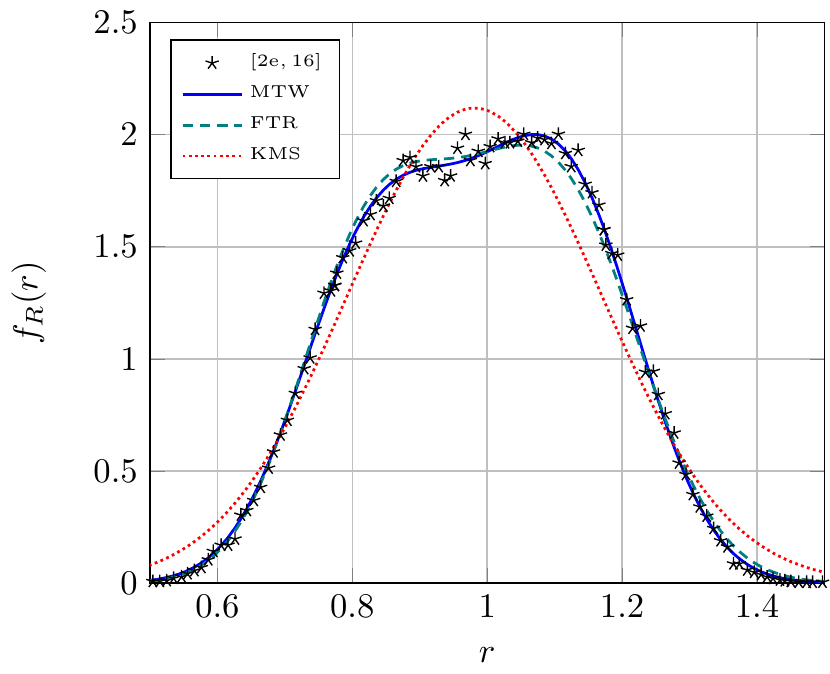}}
    \subfigure[Case F]{\includegraphics[width=0.32\textwidth]{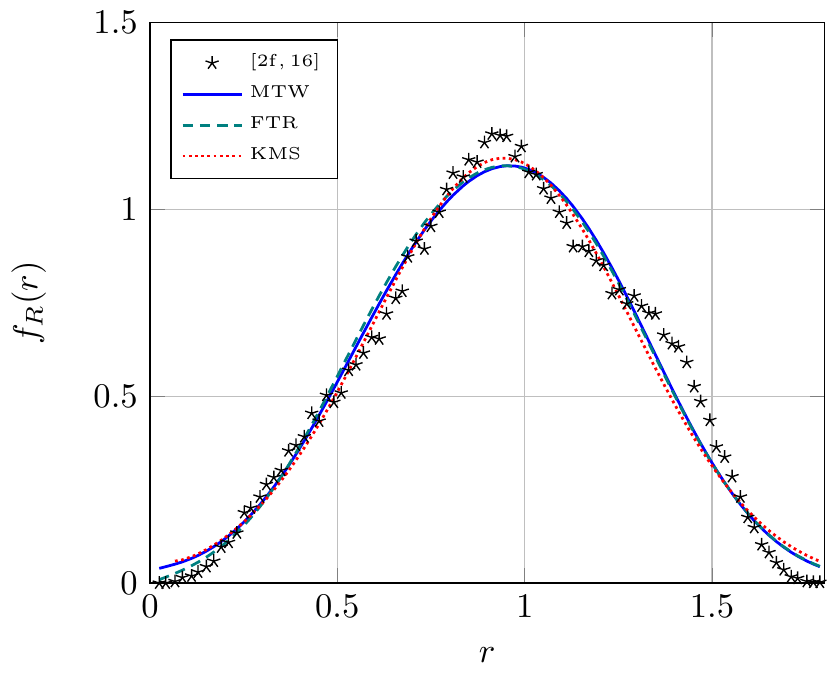}}
    \caption{\textcolor{black}{Empirical vs. theoretical PDFs for the scenarios listed in Table \ref{Tablefit}.}}
    \label{fig:fitting}
\end{figure*}

\section{Applications to wireless communication systems}

The analytical results obtained in the previous sections for the MTW fading model are now used to derive  expressions 
for different performance metrics in wireless communication systems.

\vspace{-2mm}
\subsection{Outage probability}
The channel capacity per unit bandwidth is known to be given by
\begin{equation}
\label{eq:155}
C = \log_2 (1+\gamma).
\end{equation}
The outage probability is defined as the probability that $C$ falls below a predefined threshold $R_s$ (rate per unit bandwidth), and is calculated as \cite[eq. (31)]{Romero17}
\begin{equation}
\label{eq:156}
P_{\rm out} = F_\gamma (2^{R_s}-1),
\end{equation}
where $F_{\gamma}( \cdot )$ is given by either (\ref{eq:019d}) or (\ref{eq:502}), while the asymptotic behavior is given in closed-form in (\ref{eq:113}).

\vspace{-2mm}
\subsection{Outage probability in interference-limited scenarios}
\label{OP}
We now evaluate the outage probability in an interference-limited scenario, considering an $M$-branch
MRC receiver and $L$ interfering signals with the same average power $P_I$. The received signal-to-interference ratio 
(SIR) can be written as	$SIR = \frac{X}{Y}$,
where, assuming that the interferers undergo Rayleigh fading, $X$ is the sum of $M$ independent random variables modeling the combiner output signal from the transmitter and $Y$ is 
the sum of $L$ independent exponential random variables with equal powers modeling the total interference power.
The outage probability is now defined as the probability that the SIR falls bellow a given 
threshold $\beta$, and can be calculated in terms of the generalized MGF of the desired signal as \cite[eq. (15)]{Romero08b}
\begin{equation}
	\begin{split}
		\label{eq:210}
		P_{o} = \sum_{r=0}^{L-1} 
		 \frac{1}{(\beta P_I)^r}\sum _{\tau(r,M)} \prod_{i=1}^M \frac{1}{r_i!} \left.\phi_{\gamma_i}^{(r_i)}(s)\right|_{s=-\frac{1
}{\beta P_I}},
	\end{split}
\end{equation}
where $\gamma_i$ represents in this case the received power of the desired signal at antenna $i$ affected by MTW fading
and $\tau(r,M)$ is defined as in Lemma \ref{teorema3}.
Thus, by using the GMGF given in (\ref{eq:023}), 
the outage probability under interference is obtained in closed-form. Note that $\overline \gamma$ in the computation of the GMGF represents 
in this case the average power per receive antenna, i.e., we identify $\overline{\gamma}=\overline{W}$.


\vspace{-2mm}
\subsection{Average energy detection probability}

The average probability of energy detection $\overline{P_d}$ of an unknown deterministic signal in the presence of noise in a 
wireless fading channel can be calculated as \cite{DIgham07}
\begin{equation}
\label{eq:201}
\overline{P_{d}} = \int_0^\infty {
Q_u\left(\sqrt{2\gamma}, \sqrt{\eta}\right) f_\gamma (\gamma) d\gamma},
\end{equation}
where $u=TW$, representing the the number of samples obtained in the energy detection, is the product of the one-side bandwidth 
$W$ and the observation time interval $T$,
which can be easily adjusted to get $u\in \mathbb{N}$, and $\eta$ is the energy detection threshold.
Leveraging the approach in \cite{Annamalai11}, the average detection probability can be given in terms of the generalized MGF as
\begin{equation}
\begin{split}
\label{eq:202}
\overline{P_{d}} & = 
\sum_{n=0}^\infty { \sum_{q=0}^{u+n-1} { \left(\frac{\eta}{2}\right)^q \frac{e^{-\eta/2}}{n! q!} \left.\phi_\gamma^{(n)}(s)
\right|_{s=-1}}}.
\end{split}
\end{equation}
The average detection probability of an energy detector in MTW fading can be obtained by plugging (\ref{eq:023}) into 
(\ref{eq:202}).
The so-called receiver operating characteristic (ROC) curve is obtained by representing $\overline{P_d}$ vs. ${P_f}$, for 
different values of $u$ and $\eta$, where $P_f$ is the false alarm probability, which is given by
\begin{equation}
\label{eq:203}
P_f= e^{-\eta/2}\sum^{u-1}_{k=0}\frac{(\eta/2)^k}{k!},
\end{equation}
that is, the detection, in the absence of signal, of noise which is erroneously considered to be signal.
On the other hand, a complementary ROC curve is obtained by representing $\overline{P_m}=1-{\overline{P_d}}$, defined as the probability of missed detection (i.e., failing to detect a signal which is present in the channel) vs. $P_f$.

A useful method to evaluate and compare the system performance for energy detection is the area under the ROC curve (AUC) \cite{Atapattu10}. The average AUC can be expressed in terms of the generalized MGF as \cite[(eq. 13)]{Olabiyi11c}
\begin{equation}
\begin{split}
\label{eq:204}
\overline{A} & =
1- \sum_{q=0}^{u-1} \sum_{n=0}^{q}
\binom{q+u-1}{q-n} \left(\frac{1}{2}\right)^{n+q+u}\frac{1}{n!}\left.\phi_\gamma^{(n)}(s)\right|_{s=-\frac{1}{2}},
\end{split}
\end{equation}
thus obtaining the average AUC for MTW fading by plugging the GMGF in (\ref{eq:023}) into (\ref{eq:204}).
These results can be extended to include multiple receive antennas performing MRC diversity for non-coherent energy detectors.
In this case, if $M$ is the number of antennas, the instantaneous combined SNR is given by $\gamma=\sum_{k=1}^M \gamma_k$, where $\gamma_k$ is the instantaneous SNR at the $k$  branch.
Assuming that the receive signals at every branch are independent, with the help of the multinomial theorem we can write
\begin{equation}
\begin{split}
\label{eq:205}
\phi_{\gamma}^{(r)}(s) & = \sum_{\tau(r,M)} {\frac{r!}{r_1! r_2! \cdots r_N!} \phi_{\gamma_1}^{(r_1)}(s) \cdots \phi_{\gamma_N}^{(r_M)}(s)},
\end{split}
\end{equation}
where $\tau(r,M)$ is defined as in Lemma \ref{teorema3}.
Thus, by introducing  (\ref{eq:205}) into (\ref{eq:202}) and (\ref{eq:204}) together with the GMGF expressions in (\ref{eq:023}), the performance of the energy detector for MRC diversity in MTW fading is obtained.


\subsection{Composite IG/MTW Fading Model}
We now show that the presented statistical characterization of the MTW fading model can be extended to incorporate shadowing. In particular, we consider the effect of shadowing to be modeled by the Inverse Gamma (IG) distribution, \textcolor{black}{which was proposed in \cite{karmeshu2007efficacy,eltoft2005rician} and} has been shown in \cite{ramirez2021composite}, thorough empirical validation, to yield a good fit to data measurements, particularly in
the cases of mild and moderate shadowing conditions.
The received power when IG shadowing and multipath fading are simultaneously considered will be written as
\begin{align}
	Q = \bar Q \mathcal{G}\mathcal{V},
\end{align}
where $\bar Q=\mathbb{E}\{Q\}$, and 
$\mathcal{G}$ and $\mathcal{V}$ are independent RVs  with normalized power.
 $\mathcal{G}$ is an IG random variable with shape parameter ${\lambda}$ representing shadowing, and
$\mathcal{V}$ represents an MTW fading with $\bar\gamma=1$ (note that $\gamma$ represents power in this case).
 We now show that the PDF, CDF, and outage probability of the composite IG/MTW channel model can be obtained in closed-form.

Using the closed-form GMGF in \eqref{eq:023} 
together with \cite[eq. 12]{ramirez2021composite},  the PDF of the received power $Q$ is obtained as
 \begin{align}
 	f_Q(q) = \frac{\bar{Q}^{\lambda}(\lambda-1)^{\lambda} }{q^{\lambda+1}\Gamma(\lambda)}
 	\left. \phi_{\gamma} ^{(\lambda)} (s) \right|_{s = \frac{(1-\lambda)\bar Q}{q}} ,\label{fz}
 \end{align}
 and the CDF of $Q$ for integer $\lambda$ is expressed as
 \begin{align}
 	F_Q(q) = \sum_{n=0}^{\lambda-1} \frac{\bar{Q}^{n}(\lambda-1)^{n} }{q^{n}\Gamma(n+1)} 
 	\left. \phi_{\gamma} ^{(n)} (s) \right|_{s = \frac{(1-\lambda)\bar Q}{q}}.\label{cdfz}
 \end{align}
 Besides, the outage probability is given by \cite{goldsmith2005wireless}
 \begin{align}\nonumber
 	P^{\rm IG/MTW}_{\rm out} (\gamma_{th})
 	& \triangleq P(\gamma_Q<\gamma_{th})
 	\\ \nonumber
 &	=F_{\gamma_Q}(\gamma_{th})
 \\ 
 &= F_Q \left(\frac{\bar Q \times\gamma_{th}}{\bar \gamma_Q}\right),
 \label{poutz}
 \end{align}
 where  $\gamma_Q ={ q \bar{\gamma}_Q }/{ \bar Q}$ is the SNR at the receiver, $\bar{\gamma}_Q =\mathbb{E}\{\gamma_Q\}$, $\gamma_{th}$ is the SNR threshold for reliable communication, and  $F_Q(q)$ is the CDF given in \eqref{cdfz}.

\section{Numerical results}
Numerical results are now presented for the proposed fading model using the derived expressions in the previous sections, including the representation
of the PDF and CDF of the SNR.
Different performance metrics under MTW fading are computed and verified by simulations, including
the outage probabilities in noise-limited and interference-limited scenarios and the 
average energy detection probability. Also, the outage probability for the composite IG/MTW model is provided.
The distribution parameters are $\left\{K, \Delta_i, \mu \right\}$ with $i=1\ldots N$. As in Section \ref{secVal}, when $N=1$ we will denote       $\Delta = \Delta_1$.

\begin{figure}[t]
	\centering 
	\includegraphics[height=7cm]{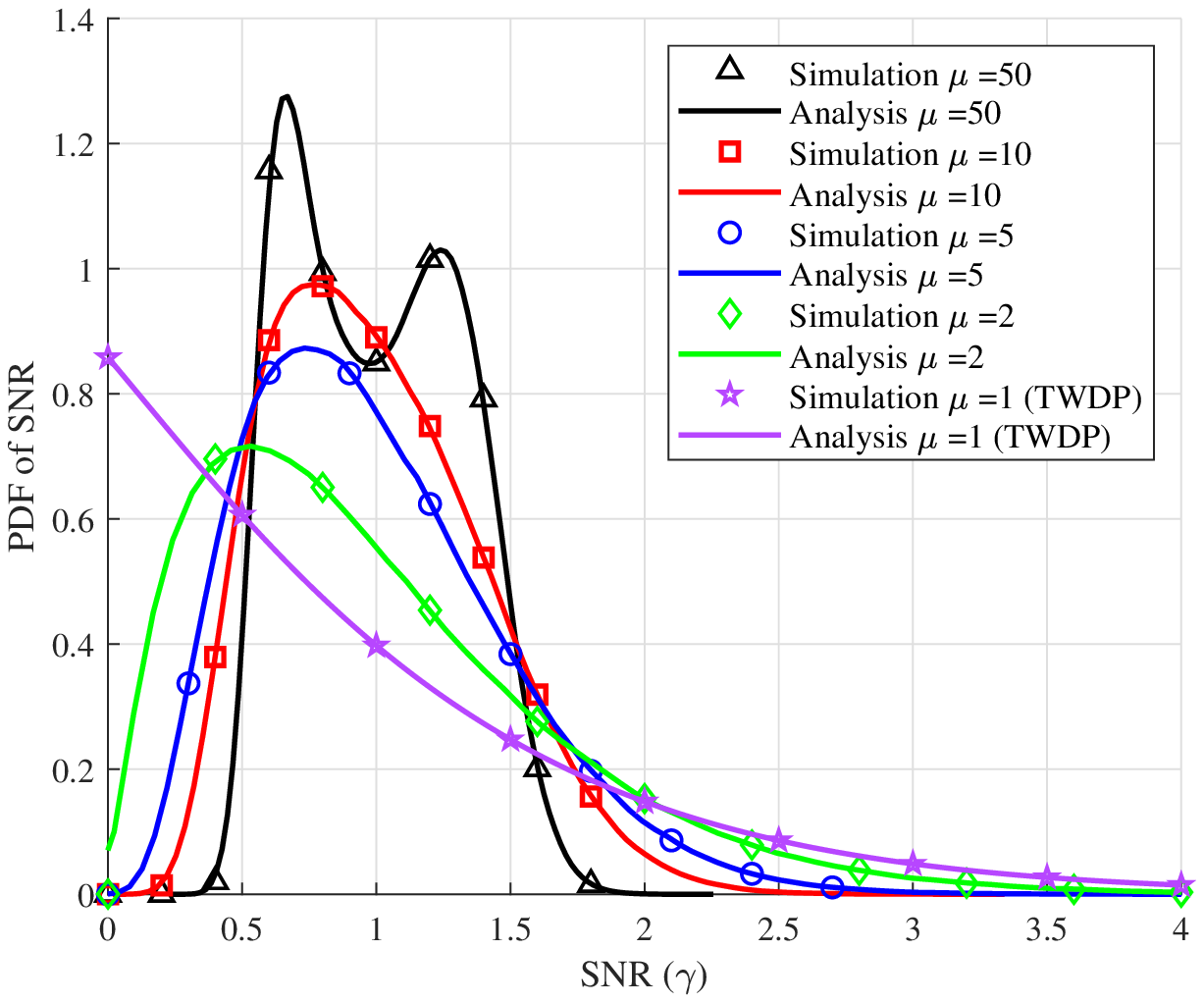}
	\caption{Analysis and simulation results for the PDF of SNR under the MTW fading model with parameters $K=1$, $\bar\gamma =  1$, $N=1$, $\Delta= 0.8$ and different values of $\mu = 2,5,10,50$.}
	\label{fig:1}
\end{figure}
\begin{figure}[t]
	\centering 
	\vspace{-2mm}
	\includegraphics[height=7cm]{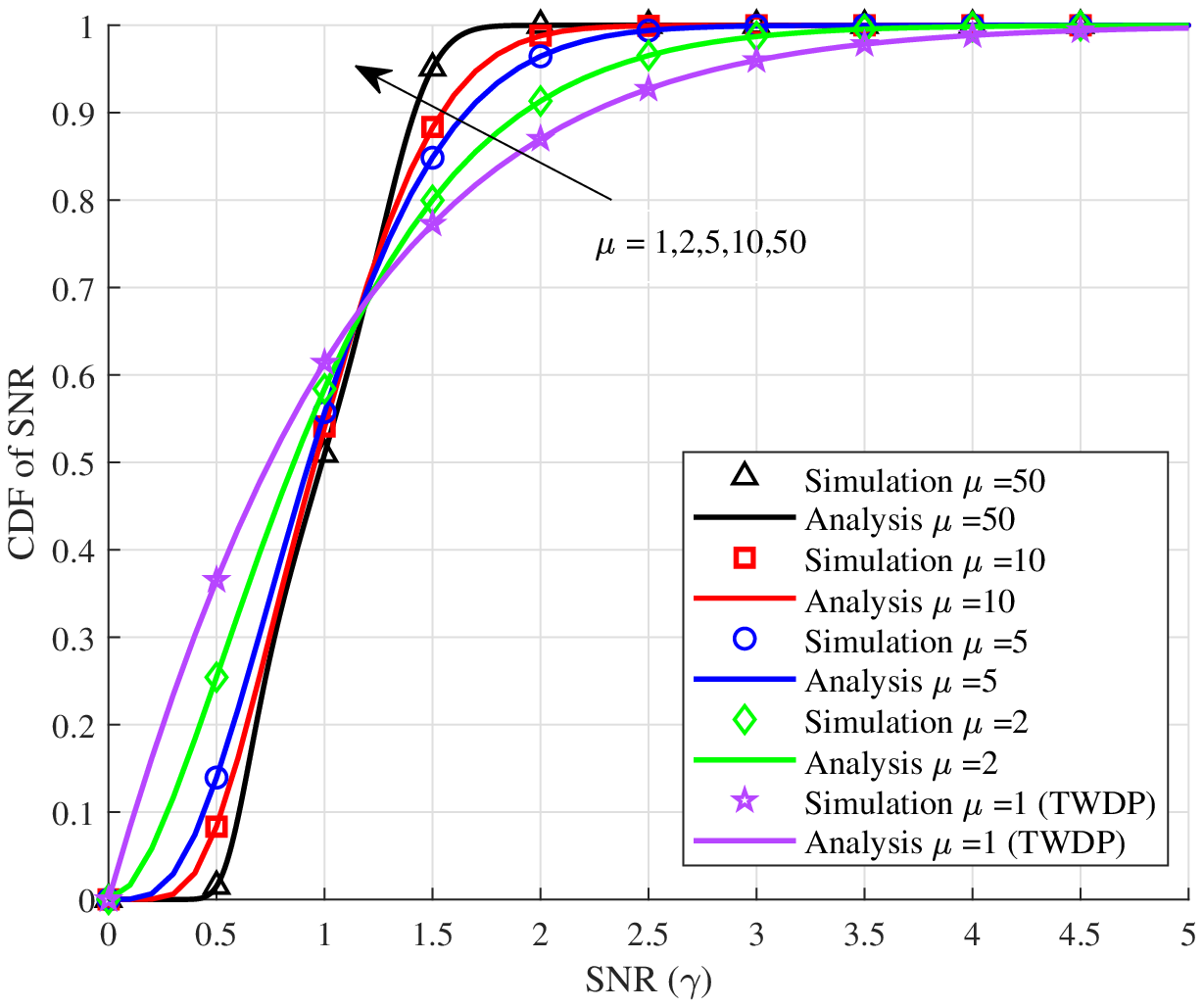}
	\caption{Analysis and simulation results for the CDF of SNR under the MTW fading model with parameters $K=1$, $\bar\gamma =  1$, $N=1$, $\Delta= 0.8$ and different values of $\mu = 2,5,10,50$. }
	\label{fig:2}
\end{figure}

\begin{figure}[t]
	\centering 
	\includegraphics[height=7cm]{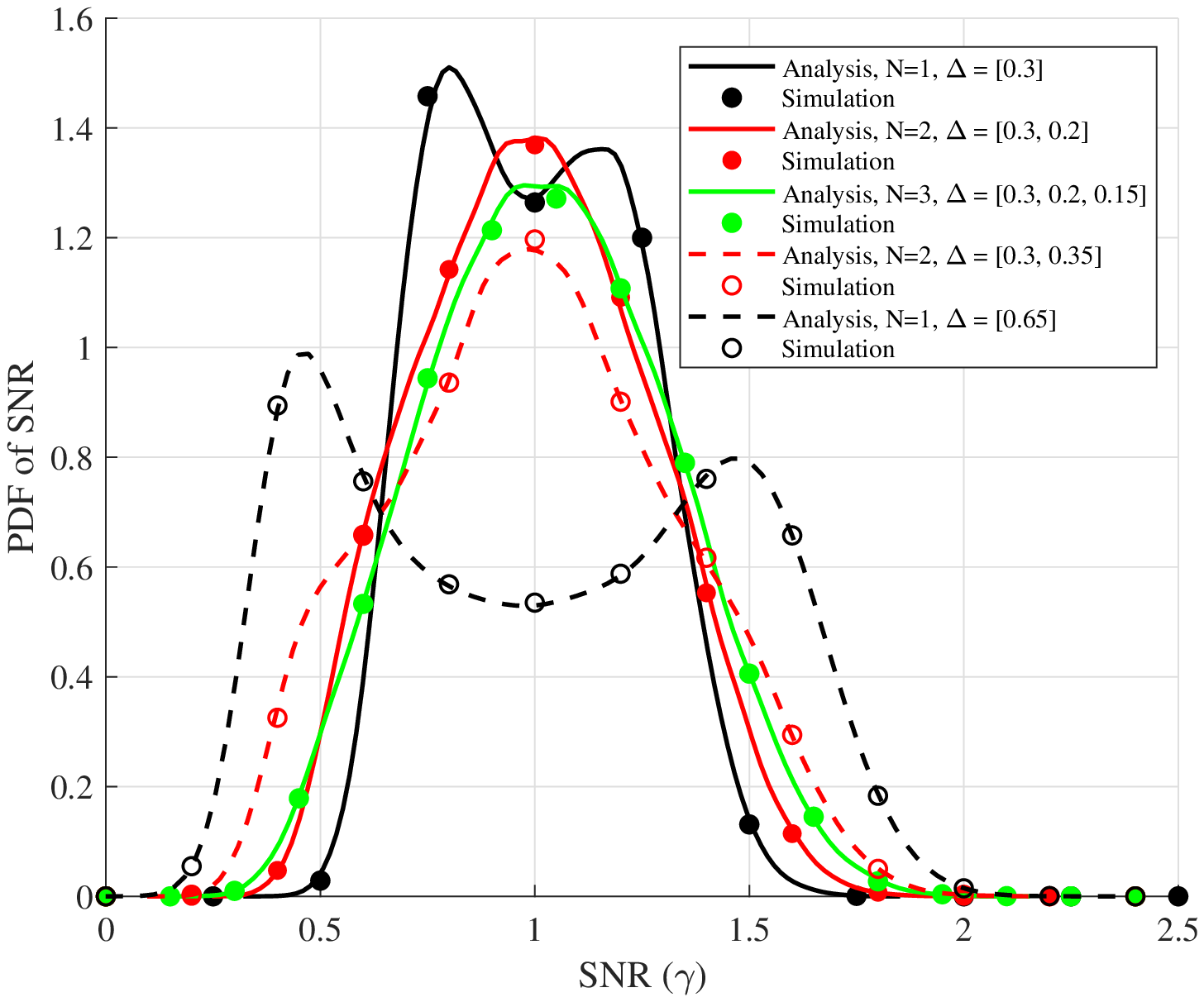}
	\caption{Analysis and simulation results for the PDF of SNR under the MTW fading model with parameters $K=15$, $\bar\gamma =  1$, $\mu= 10$ and different values of $N$ and $\Delta  = \left[ {\Delta _1 , \ldots ,\Delta _N } \right]$.}
	\label{pdf}
\end{figure}

\begin{figure}[t]
	\centering 
	\includegraphics[height=7cm]{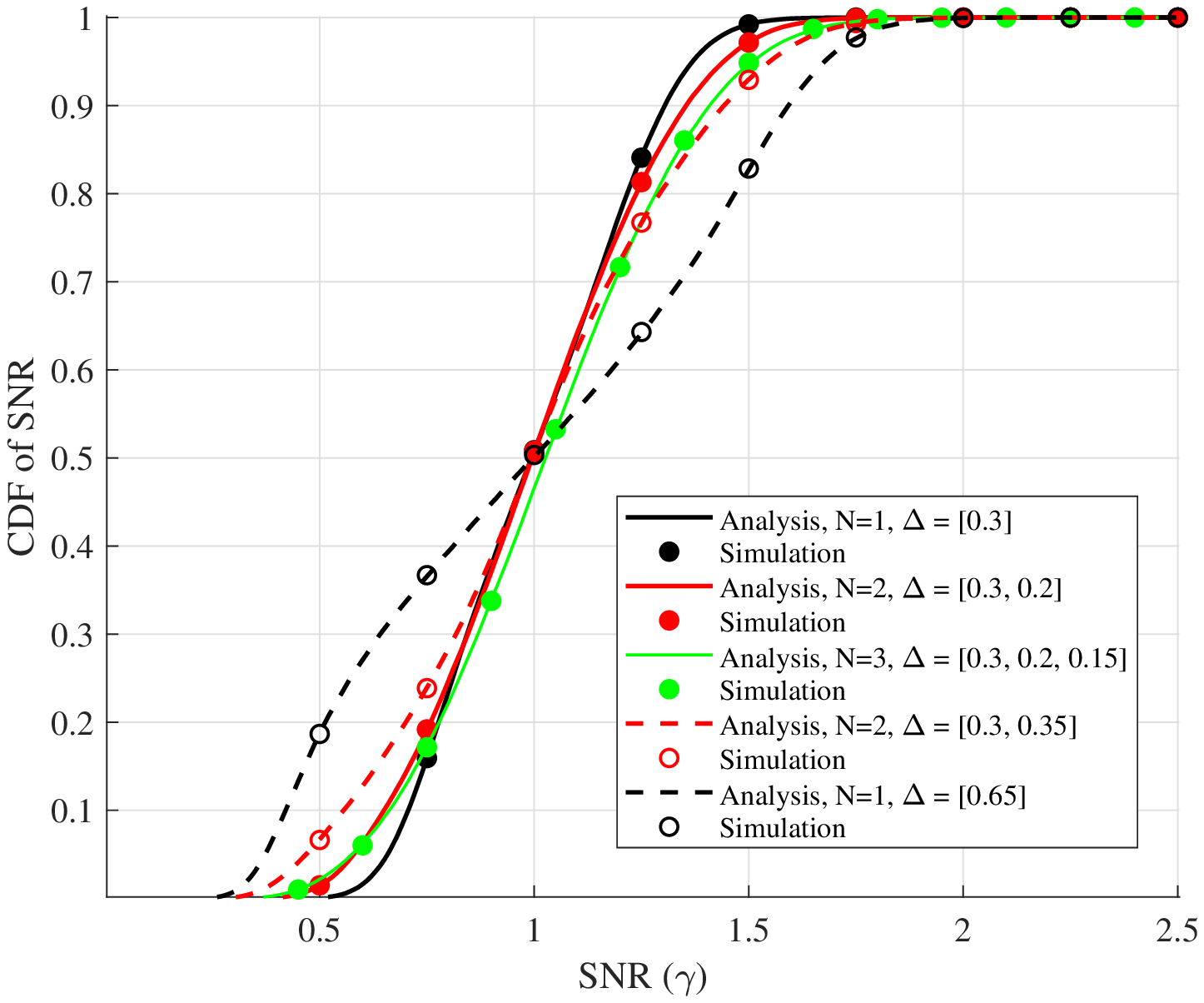}
	\caption{Analysis and simulation results for the CDF of SNR under the MTW fading model with parameters $K=15$, $\bar\gamma =  1$, $\mu= 10$ and different values of $N$ and $\Delta  = \left[ {\Delta _1 , \ldots ,\Delta _N } \right]$.}
	\label{cdf}
\end{figure}

\begin{figure}[t]
	\centering
	\includegraphics[height=7cm]{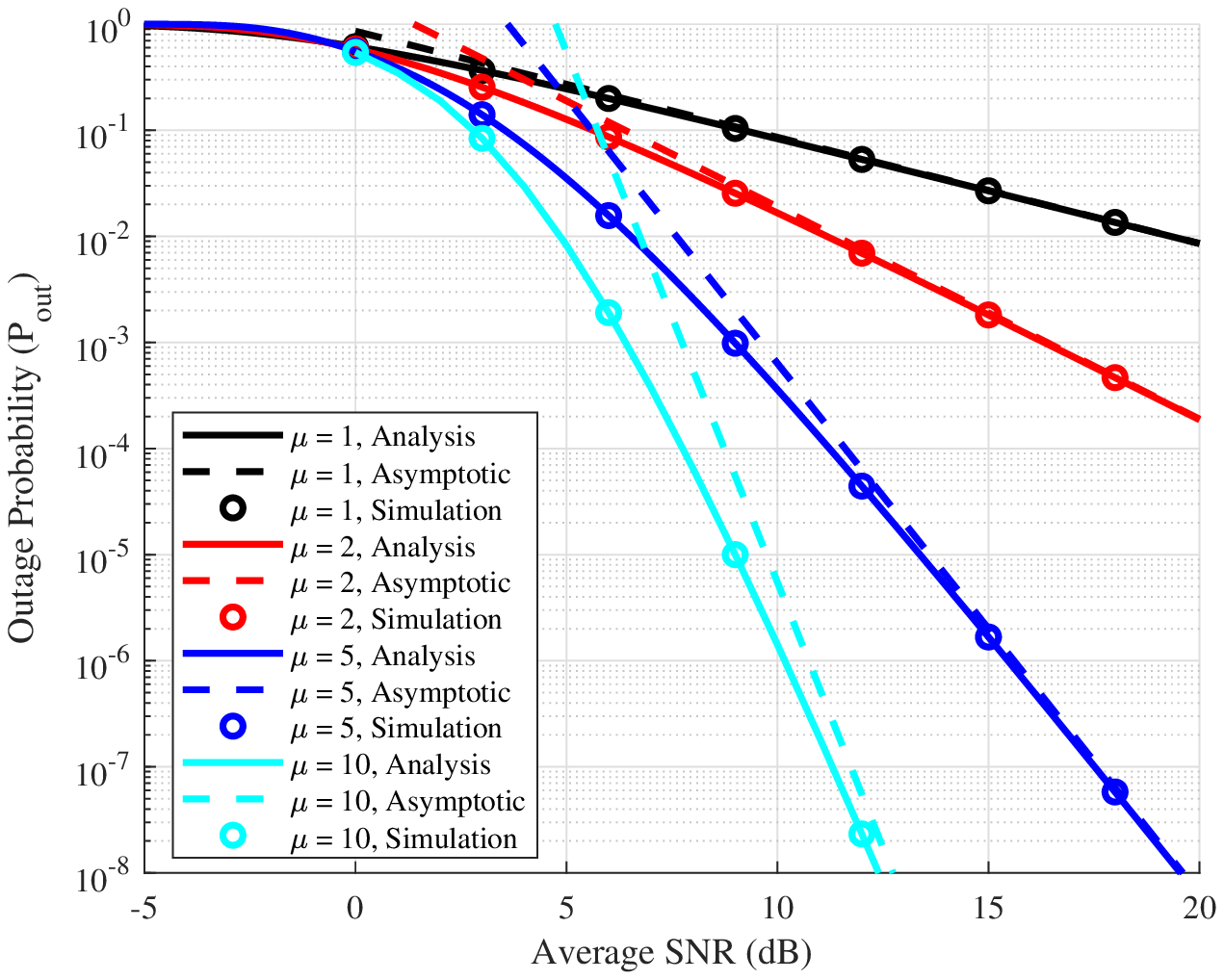}
	\caption{Outage probability $P_{out}$ vs. average SNR for MTW fading, asymptotic behavior in high SNR considering different number of clusters ($\mu=1,2,5,10$) and parameters $K=1$, $N=1$, $\Delta =0.8$, and $R_s=1$. }
	\label{fig:pout}
\end{figure}

In Figs. \ref{fig:1} and \ref{fig:2}, the PDF and CDF, respectively, of the MTW fading model are plotted for different values 
of $\mu$, including the case $\mu=1$, which corresponds to the TWDP fading model. 
The numerical results are obtained indistinctly either from \eqref{eq:019} or \eqref{eq:501} for the PDF, and \eqref{eq:019d} or \eqref{eq:502} for the CDF, and are verified by Monte-Carlo simulations, showing an excellent match. The obtained results show that when 40 summation terms are considered in \eqref{eq:501} or \eqref{eq:502} the figures cannot be distinguished from those obtained with the integral representations of the probability functions. 
It is clear from Fig. \ref{fig:1} that parameter $\mu$ has a relevant impact on the shape of the probability distribution, even to the point that,
for high values of this parameter ($\mu=50$) the distribution shows a bimodal behavior, i.e., two local maxima appear in the PDF,
while for lower values ($\mu=2,5,10$) only one local (and therefore global) maximum appears. 
\textcolor{black}{In these figures, parameters $K=1$, $\bar\gamma =  1$, $\Delta= 0.8$ are fixed. A large value of $\Delta= 0.8$ implies that the specular power is divided almost equally between the two components.  Also, because of the relationship $\bar\gamma=(E_s / N_0) 2{\sigma ^2}\mu (1+K)$, as $\mu$ increases, necessarily the diffuse power per cluster, $2{\sigma ^2}$, decreases, resulting in the model distribution being dominated by the specular power splitting in the two components, yielding this bimodal behavior.}
It is also interesting to note that, for the selected parameter values, the case $\mu=1$, i.e., TWDP fading, 
yields a monotonic decreasing PDF, with a maximum at the origin. 
\textcolor{black}{In the CDF curves depicted in Fig. \ref{fig:2}, the bimodal behavior is manifested in the appearance of multiple inflection points, 
although they can hardly be observed with the naked eye. On the other hand, it can be clearly seen that the distribution becomes more concentrated as $\mu$ increases, which manifests in a steepest CDF, and which is consistent with the decrease of the normalized variance (i.e., the AoF) given in \eqref{eq:111} for large values of $\mu$.}
Therefore, parameter $\mu$ alone gives the model a great flexibility to fit to different propagation conditions.

\textcolor{black}{
Figs. \ref{pdf} and \ref{cdf} illustrate the PDF and CDF, respectively, of the SNR for different values of $N$, which represents the number of clusters containing two specular components, maintaining fixed parameters $\mu=10$ and $K=15$. In the figures legends, $\Delta$ is defined as a vector with $N$ components, i.e., $\Delta  = \left[ {\Delta _1 , \ldots ,\Delta _N } \right]$. It must be remarked that the model is oblivious of the order of the elements of such vector. In both the PDF and CDF curves, we observe that, as $N$ increases the degrees of freedom of the MTW model, very different shapes of both the PDF and CDF are obtained, which permits to model many different wireless environments.  It is also clear that even for $N=1$, the $\Delta$ parameter have a very significant impact on the model statistics.
In the remaining of this section we will assume $N=1$.}


\begin{figure}[t]
	\centering
	\includegraphics[height=7cm]{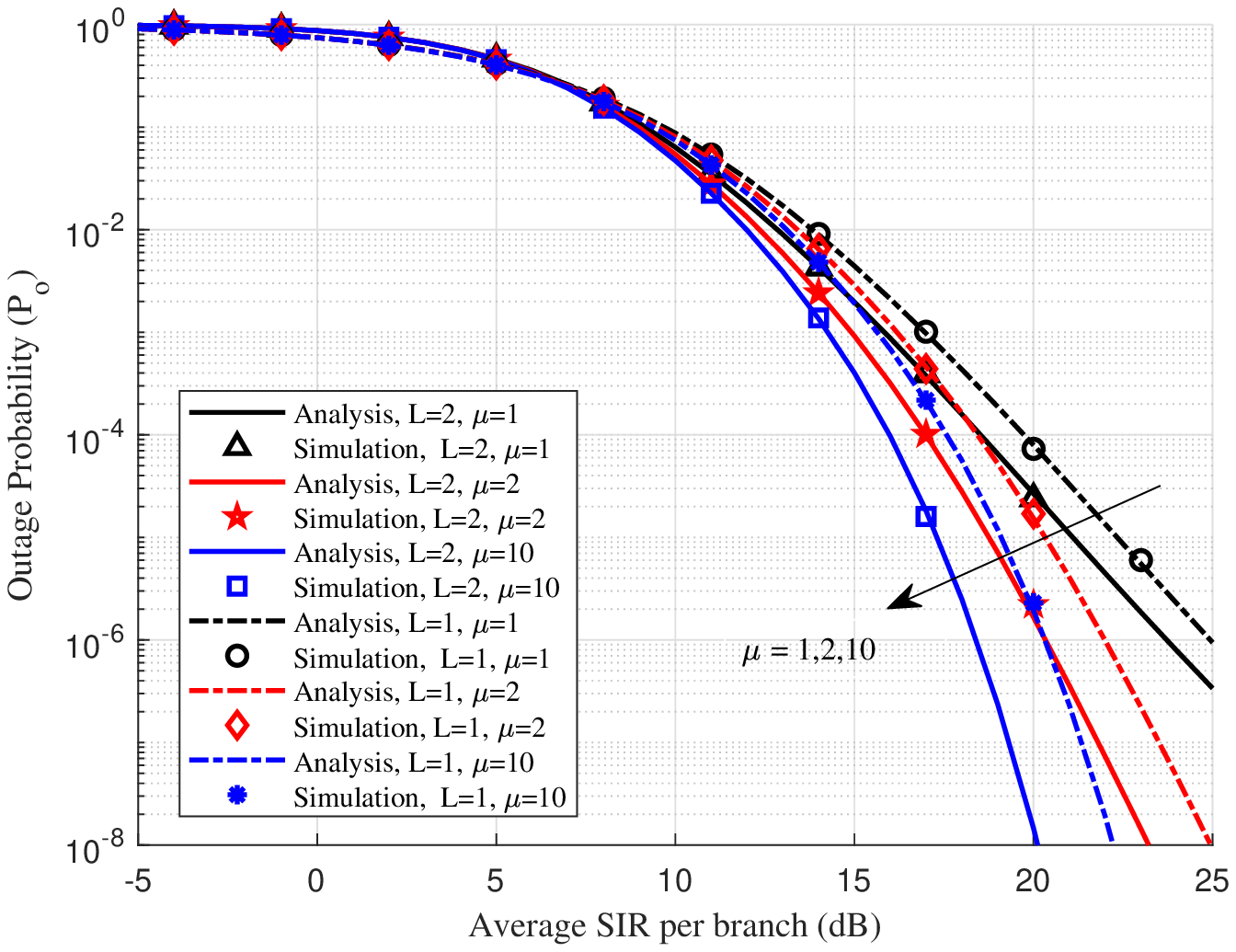}
	\caption{Outage probability $P_o$ vs. average SIR per branch for MTW fading considering different 
number of clusters ($\mu=1,2,10$) and interferers ($L=1,2$), and parameters $K=10$, $N=1$, $\Delta =0.8$, $M=3$, and $\beta =10$. }
	\label{fig:4}
\end{figure}

\begin{figure}[t]
	\centering
	\includegraphics[height=7cm]{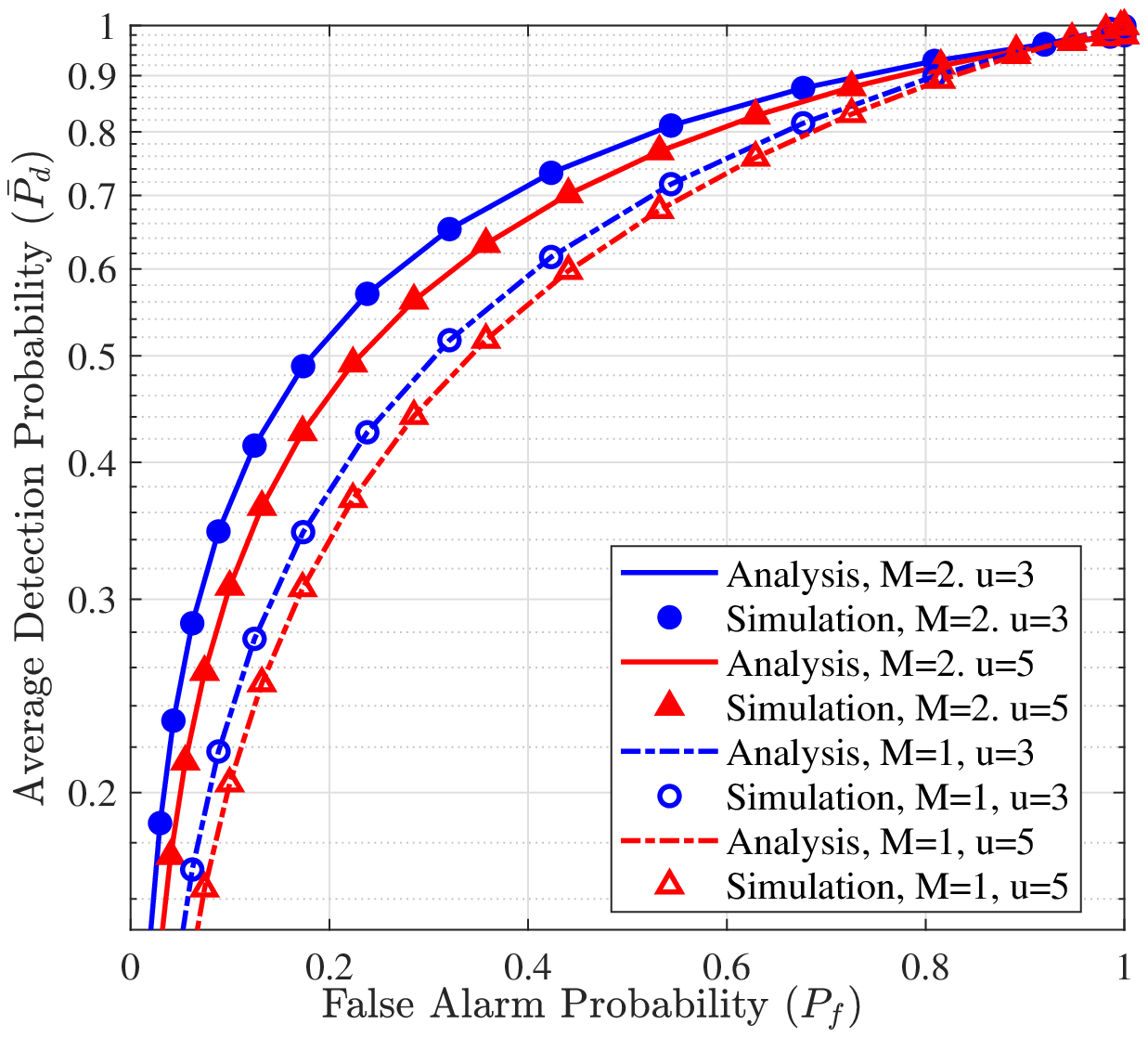}
	\caption{ROC curve (average $\overline {P_d}$ vs. average ${P_f}$ ) under MTW fading for different numbers of receive 
		diversity branches $M$ and number of samples $u$, when $K=10$ and $\mu = 5$, $N=1$, $\Delta=0.3$ and  $\bar\gamma=1$.  }
	\label{fig:5}
\end{figure}
\begin{figure}[t]
	\centering
	\vspace{-0.75mm}
	\includegraphics[height=7cm]{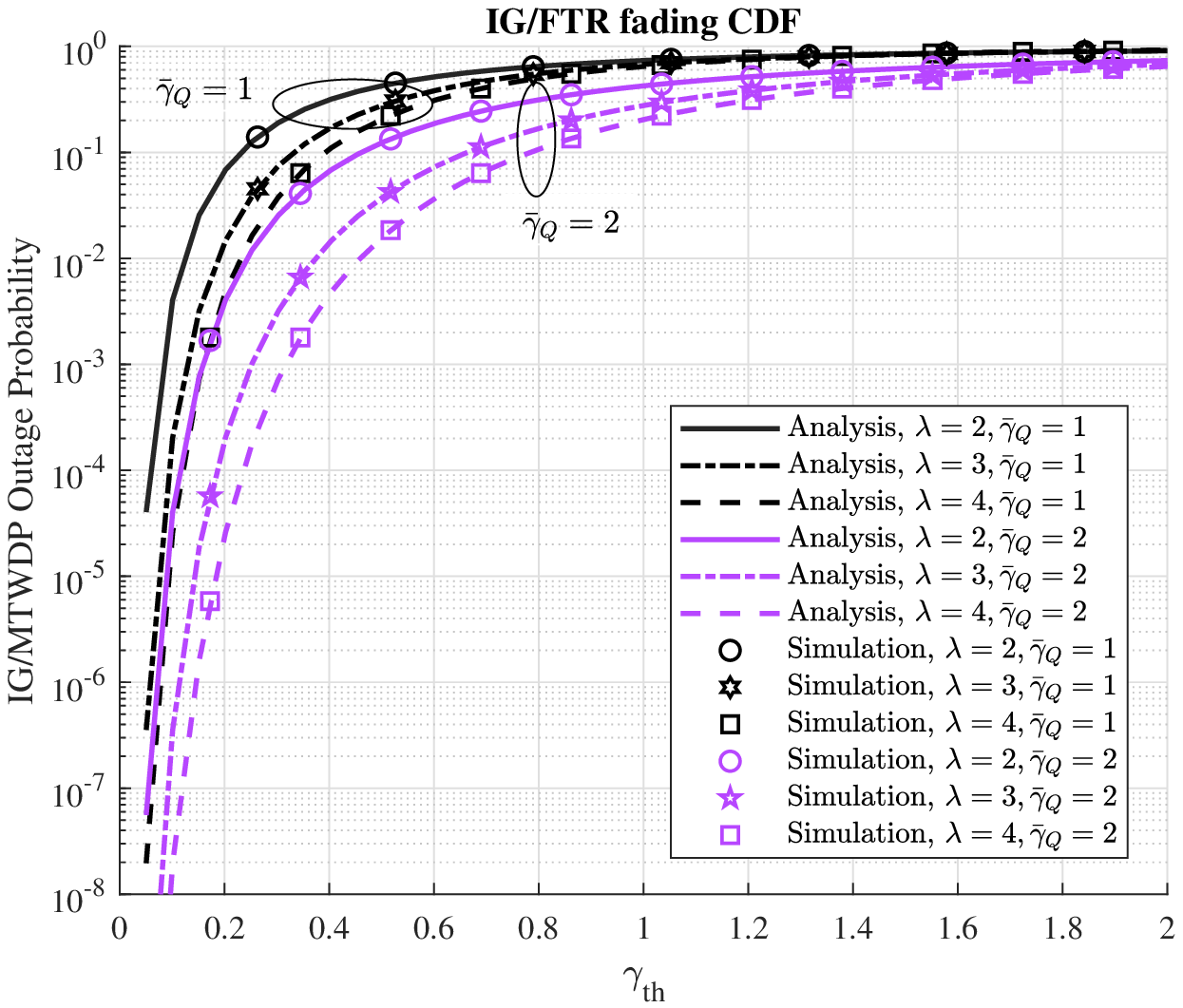}
	\caption{Outage probability of the composite IG/MTW model vs. SNR threshold ($\gamma_{th}$) with parameters $\mu=5$, $K=10$, $N=1$, $
\Delta=0.3$, $\bar\gamma=1 $, and $\bar Q=1$.}
	\label{fig:8}
\end{figure}

\textcolor{black}{Fig. \ref{fig:pout}  presents the outage probability vs. average SNR without interference, obtained from \eqref{eq:156}}, where the asymptotic behavior is also shown and
different numbers of clusters $\mu=1,2,5,10$ are considered. It can be observed that the outage 
probability decreases by raising the number of clusters $\mu$.
The asymptotic curves clearly show the effect of the diversity order (higher slope in the high SNR regime as $\mu$ increases).

\textcolor{black}{Fig. \ref{fig:4} depicts the outage probability, obtained from \eqref{eq:210} together with \eqref{eq:023}, vs. the average SIR per branch} , $\overline{W}/(L\cdot P_I)$, in 
an interference-limited scenario with an $M$-branch MRC receiver under MTW 
fading and $L$ interferers experiencing Rayleigh fading with equal average power $P_I$.
The analytical results are verified by  Monte-Carlo simulations, 
considering parameter values $\mu=1,2,10$ and $L=1,2$.
It can be observed that the outage probability decreases by increasing $\mu$, i.e., it is 
more beneficial that the signal power is distributed among different clusters.
\textcolor{black}{Also, for a given fixed average SIR, increasing the number of interferes reduces the probability of outage. This is due to the fact that the total (fixed) average  interference power is divided among the individual interferers. Therefore the total interference tends to average, reducing its variance, which yields a lower outage probability.}

Fig. \ref{fig:5} presents ROC curves ($\overline{P_d}$ vs. ${P_f}$) to analyze the effect on an energy detector of parameter $u$
 (number of samples)
and the number of receive antennas at the MRC receiver when the channel undergoes MTW fading. The average probability of energy 
detection ($\overline{P_d}$) and 
false alarm probability ($P_f$) are given in \eqref{eq:202} and \eqref{eq:203}, respectively.
It can be observed that the ROC curves rise noticeably when the number of MRC branches increases and, for a given number of 
branches in the MRC receiver, lower $u$ shows a better result. Thus, increasing the number of samples $u$ 
decreases the AUC for a given SNR per branch. 
%




Finally, Fig. \ref{fig:8} presents numerical results of the outage probability for the IG/MTW channel model.
 The outage probability in \eqref{poutz} is plotted as a function of the SNR threshold for different values of the IG parameter 
$\lambda=2,3,4$ and for average SNR $\bar{\gamma}_Q=1,2$. 
As expected, the outage probability increases by rising the SNR threshold and, for a given threshold, increasing 
the average SNR reduces the outage probability.
Also,  low $\gamma_{th}$ and high values of the shape parameter of the IG distribution $\lambda$ yields a lower outage probability.



%
%

\section{Conclusions}

We have presented the Multi-cluster Two-Wave (MTW) fading model, by defining its physical model and deriving its chief probability 
functions. The unique features of the MTW model allow to simultaneously control both the diversity order and the distribution 
bimodality, encompassing   the well-established $\kappa$-$\mu$ and TWDP fading models as special cases. 
The derived expressions for the MTW model, which are 
given in closed-form for the case of Laplace-domain statistics, and in finite-range integral form or in series form (as a mixture of Gamma distributions) for the PDF and CDF, 
can be directly applied for performance analysis purposes and have been consequently exemplified. \textcolor{black}{The proposed model has been validated by providing fitting results to experimental measurements, and has been shown to outperform other generalized fading models such as FTR and KMS in some environments in the sub-THz band, which is expected to be used in 6G wireless communications networks.}

The MTW fading model can be further extended to include fluctuations of the specular components \cite{vega2022multi}, thus connecting with the FTR and the KMS fading models. However, it must be noted that the formulation in \cite{vega2022multi} only considers one cluster with two specular components, while in the MTW model here proposed there is an arbitrary number of clusters $N$ with this characteristic. Thus \cite{vega2022multi} only connects with the MTW fading for $N=1$ and, even in that case, as a limit by letting one of the parameters grow to infinity, while the derivations here presented have compact form.

\appendices
\section{Proof of Lemma 2} \label{apendiceA}
Taking into account the power series expansion of the modified Bessel functions $I_{\nu}$ given in \cite[(8.445)]{Gradshteyn00}, (\ref{eq:018}) can be written as
\begin{equation}
\label{eq:B01}
\begin{split}
  f_{\gamma _\theta  } (x) &= \frac{{\mu \left( {1 + K} \right)^{\frac{{\mu  + 1}}
{2}} }}
{{\overline \gamma  e^{\mu \kappa _\theta  } \;}}\left( {\frac{x}
{{\overline \gamma  \kappa _\theta  }}} \right)^{\frac{{\mu  - 1}}
{2}} e^{ - \frac{{\mu \left( {1 + K} \right)}}
{{\overline \gamma  }}x}  \hfill \\ &
   \times \sum\limits_{k = 0}^\infty  {\frac{1}
{{k!\Gamma \left( {\mu  + k} \right)}}} \left( {\frac{{2\mu \sqrt {\frac{{\kappa _\theta  \left( {1 + K} \right)}}
{{\overline \gamma  \;}}x} }}
{2}} \right)^{\mu  - 1 + 2k},
\end{split}
\end{equation}
which, by considering (\ref{eq:016}) and using the binomial and the multinomial theorems, can be rewritten after some manipulation as
\begin{equation}
\label{eq:B02}
\begin{split}
  f_{\gamma _\theta  } (x) &= e^{ - \mu K} e^{ - \frac{{\mu \left( {1 + K} \right)}}
{{\overline \gamma  }}x} \sum\limits_{k = 0}^\infty  {K^k \mu ^{\mu  + 2k} }  \hfill \\&
   \times \left( {\frac{{1 + K}}
{{\overline \gamma  }}} \right)^{\mu  + k} \frac{{x^{\mu  + k - 1} }}
{{k!\Gamma \left( {\mu  + k} \right)}}\sum\limits_{r = 0}^k \binom{k}{r} \hfill \\&
   \times \sum\limits_{\tau (r,N)} {\frac{{r!}}
{{r_1 !r_2 ! \cdots r_N !}}} \prod\limits_{i = 1}^N {\left( {\Delta _i \cos \theta _i } \right)^{r_i } }  \hfill \\&
   \times \left[ {\prod\limits_{i = 1}^N {e^{ - \mu K\Delta _i \cos \theta _i } } } \right].
\end{split}
\end{equation}
By averaging over $\theta_1,\theta_2,\ldots,\theta_N$, the unconditional PDF of the SNR can be expressed as
\begin{equation}
\label{eq:B03}
\begin{split}
  f_\gamma  (x) &= e^{ - \mu K} e^{ - \frac{{\mu \left( {1 + K} \right)}}
{{\overline \gamma  }}x} \sum\limits_{k = 0}^\infty  {K^k \mu ^{\mu  + 2k} \left( {\frac{{1 + K}}
{{\overline \gamma  }}} \right)^{\mu  + k} }  \hfill \\&
   \times \frac{{x^{\mu  + k - 1} }}
{{k!\Gamma \left( {\mu  + k} \right)}}\sum\limits_{r = 0}^k \binom{k}{r} r!  \sum\limits_{\tau (r,N)} \prod\limits_{i = 1}^N \frac{{\Delta _i^{r_i } }}
{{r_i !}}\hfill \\&
   \times\frac{1}
{\pi }\int_0^\pi  {e^{ - \mu K\Delta _i \cos \theta _i } \left( {\cos \theta _i } \right)^{r_i } d\theta _i }  .
\end{split}
\end{equation}
With the help of \cite[eq. (14)]{ermolova2016capacity}, the integral in (\ref{eq:B03}) can be solved in closed-form as
\begin{equation}
\label{eq:B04}
\begin{split}
\frac{1}
{\pi } \int_{0}^\pi { \exp \left(\alpha \cos\theta \right) (\cos \theta)^{m} d\theta }
= \frac{1}{2^m} \sum_{l=0}^m \binom{m}{l} I_{2l-m}(\alpha).
\end{split}
\end{equation}
Then, using
(\ref{eq:B04}) in (\ref{eq:B03}), (\ref{eq:501}) is obtained. Finally, (\ref{eq:502}) is obtained by integrating (\ref{eq:501}).

\section{Proof of Lemma 3} \label{apendiceB}
For a particular realization of $\theta_1,\theta_2,\ldots,\theta_N$ the MTW model collapses to the $\kappa$-$\mu$ model, whose GMGF was provided in \cite[Table II]{ramirez2021composite} as
\begin{equation}
\label{eq:A01}
\begin{split}
\phi _{\kappa  - \mu }^{(n)} (s) & = \frac{{\overline {\gamma _\theta  } ^n \Gamma \left( {\mu  + n} \right)\mu ^\mu  \left( {1 + \kappa _\theta  } \right)^\mu  e^{ - \mu \kappa _\theta  } }} 
{{\Gamma \left( \mu  \right)\left( {\mu \left( {1 + \kappa _\theta  } \right) - \overline {\gamma _\theta  } s} \right)^{\mu  + n} }}
\\& \times
_1 F_1 \left( {\mu  + n:\mu ;\frac{{\mu ^2 \kappa _\theta  \left( {1 + \kappa _\theta  } \right)}}
{{\mu \left( {1 + \kappa _\theta  } \right) - \overline {\gamma _\theta  } s}}} \right),
\end{split}
\end{equation}
which, from (\ref{eq:015b}) and (\ref{eq:016}), and also considering \cite{Wolfram} and the multinomial theorem, can be rewritten, after some manipulation, as
\begin{equation}
\label{eq:A02}
\begin{split}
  \phi _{_{\gamma \left| \theta  \right.} }^{(n)} (s) & = \overline \gamma  ^n \Gamma \left( {\mu  + n} \right)e^{\frac{{\mu K\overline \gamma  s}}
{{\mu \left( {1 + K} \right) - \overline \gamma  s}}} \sum\limits_{q = 0}^n  \binom{n}{q} \frac{{\left( {\mu K} \right)^q }}
{{\Gamma (\mu  + q)}} \\& \times
\frac{{\left( {\mu \left( {1 + K} \right)} \right)^{q + \mu } }}
{{\left( {\mu \left( {1 + K} \right) - \overline \gamma  s} \right)^{q + \mu  + n} }} \hfill \left[\prod\limits_{i = 1}^N {e^{\frac{{\mu K\Delta _i \overline \gamma  s}}
{{\mu \left( {1 + K} \right) - \overline \gamma  s}}\cos \theta _i } }\right] \\& \times \sum\limits_{r = 0}^q  \binom{q}{r} \sum\limits_{\tau (r,N)} {\frac{{r!}}
{{r_1 !r_2 ! \cdots r_N !}}\prod\limits_{i = 1}^N {\left( {\Delta _i \cos \theta _i } \right)^{r_i } } }.  
\end{split}
\end{equation}
By averaging over $\theta_1,\theta_2,\ldots,\theta_N$, the unconditional GMGF can be expressed as
\begin{equation}
\label{eq:A004}
\begin{split}
\phi _{_\gamma  }^{(n)}  & (s)  = \overline \gamma  ^n \Gamma \left( {\mu  + n} \right)e^{\frac{{\mu K\overline \gamma  s}}
{{\mu \left( {1 + K} \right) - \overline \gamma  s}}} \sum\limits_{q = 0}^n \binom{n}{q} \frac{{\left( {\mu K} \right)^q }}
{{\Gamma (\mu  + q)}} \\& \times \frac{{\left( {\mu \left( {1 + K} \right)} \right)^{q + \mu } }}
{{\left( {\mu \left( {1 + K} \right) - \overline \gamma  s} \right)^{q + \mu  + n} }} 
 \sum\limits_{r = 0}^q \binom{q}{r} r!\sum\limits_{\tau (r,N)} \prod\limits_{i = 1}^N \frac{{\Delta _i^{r_i } }}
{{r_i !}} \\& \times \frac{1}
{\pi }   \int_0^\pi  {e^{\frac{{\mu K\Delta _i \overline \gamma  s}}
{{\mu \left( {1 + K} \right) - \overline \gamma  s}}\cos \theta _i } \left( {\cos \theta _i } \right)^{r_i } d\theta _i }  .
\end{split}
\end{equation}
The integral in (\ref{eq:A004}) can be solved in closed-form as given in (\ref{eq:B04}), yielding (\ref{eq:023}).


\renewcommand{\baselinestretch}{1}
\normalsize

{\LARGE
\bibliographystyle{IEEEtran}
\bibliography{IEEEabrv,Mi_bibliografia}
}

\end{document}